\documentclass{pinchcr}   % Comment the above and uncomment this
\newcommand{\beq}{\begin{equation}}
\newcommand{\eeq}{\end{equation}}
\newcommand{\bmat}{\begin{displaymath}}
\newcommand{\emat}{\end{displaymath}}
\newcommand{\eq}[1]{Eq.~(\ref{#1})}

\usepackage{graphicx}        % standard LaTeX graphics tool
\usepackage{amsmath}

\def\thesection{\arabic{section}}
\def\thefigure{\arabic{section}.\arabic{figure}}

\title{Growing length scales in aging systems}

\author{Federico Corberi}
\affiliation{Dipartimento di Matematica ed Informatica, Universit\`a
di Salerno, via Ponte don Melillo, 84084 Fisciano (SA), Italy}

\author{Leticia F. Cugliandolo}
\affiliation{Universit\'e Pierre et Marie Curie - Paris VI,
Laboratoire de Physique Th\'eorique et Hautes Energies, 4 Place
Jussieu, Tour 13, 5\`eme \'etage, 75252 Paris Cedex 05, France.}

\author{Hajime Yoshino}
\affiliation{Department of Earth and Space Science, Faculty of
Science, Osaka University, Toyonaka 560-0043, Japan.}

\begin{document}

\maketitle
%\date{April 2, 2010}

\preface{We summarize studies of growing lengths in different aging
systems. The article is structured as follows. We recall the
definition of a number of observables, typically correlations and
susceptibilities, that give access to dynamic and static correlation
lengths. We use a growing length perspective to review three out of
equilibrium cases: domain growth phenomena; the evolution of
Edwards-Wilkinson and Kardar-Parisi-Zhang manifolds and other directed
elastic manifolds in random media; spin and structural glasses in
relaxation and under an external drive. Finally, we briefly report on
a mechanism for dynamic fluctuations in aging systems that is based on
a time-reparametrization invariance scenario and may be at the origin
of the dynamic growing length in glassy materials.}

\maintext

\chapter*{}

\section{Introduction}
\label{sec:intro}

For a long time, and somewhat paradoxically, the majority of
theoretical physicists interested in glasses were reluctant to study
these systems in the trully glassy regime, say below the glass
temperature $T_g$ or above the glass density $\rho_g$.  One of the
reasons to resist entering the glassy regime was the lack of insight
on which questions to ask and, more concretely, which quantities to
measure in out of equilibrium conditions.

The situation changed dramatically around 20 years ago when it was
accepted that glassy systems do not freeze out below $T_{g}$ but they
just continue to relax at a slower and slower rate as time goes by --
the {\it aging} phenomenon.  Although this fact was well-known
experimentally and to a certain extent also numerically it was not
fully assimilated theoretically. The solution to a number of simple
models [namely, droplet models for disordered
systems~\shortcite{Fisher-Huse}, random walks among traps in phase
space~\shortcite{Bouchaud}, and mean-field spin systems with disordered
interactions~\shortcite{Cuku1,Cuku2}] demonstrated the possibility of capturing
aging phenomena analytically. These solutions also provided a
guideline as to which are the simplest preparation protocol and the
most relevant observables to focus on dynamically.  More importantly,
these studies also opened the way to confront glassy dynamics to the
behaviour of other macroscopic non-equilibrium systems considered to
be simpler such as coarsening phenomena~\shortcite{Bray94} or
the motion of manifolds in different kinds of embedding volumes~\shortcite{manifold}.
Experimental and numerical data on theoretically motivated
measurements in various physical systems have
rapidly accumulated and some of these results are
summarized in other chapters in this book.

As for the measuring protocols, 
in short, one first chooses the way in which the system is taken into
the glassy regime and defines the origin of the time axis
accordingly. In the simplest setting, the `initial time' $t=0$, is the
instant when the system is suddenly prepared in an out-of-equilibrium
condition.  For instance, this is achieved by performing a rapid
temperature quench from a high temperature above $T_{g}$ down to a
target value $T< T_g$ but other routes to the glassy regime, changing other control 
parameters, are also feasible. As will be discussed below, the physical
properties of the system become functions of the waiting time $t_w$,
the time spent after the preparation of the initial state, stationarity is 
lost and the
relaxation time increases with $t_w$.  The basic idea to analyze such
an aging relaxation is to introduce a laboratory time scale -- the
waiting-time -- that can be controlled at will when the intrinsic time
scale -- the equilibrium relaxation time -- goes beyond accessible
times.

\begin{figure}[h]
\vspace{1cm}
\begin{center}
%\centerline{
%\hspace{-2.25cm}
%\input{sketch-corr.pslatex}
%\hspace{-3cm}
%\input{sketch-chi.pslatex}
%}
\includegraphics[width=0.9\textwidth]{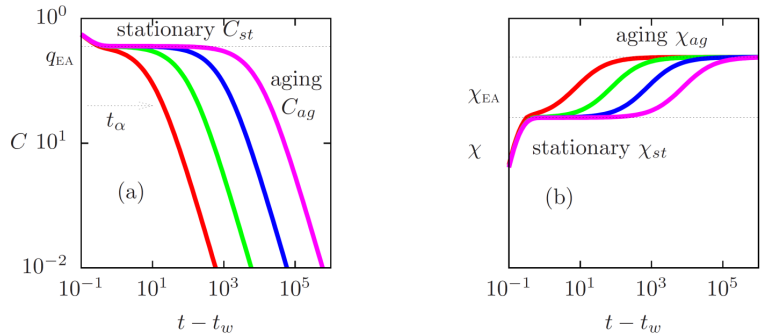}
\end{center}
%\vspace{-1cm}
\caption{Typical behaviour of macroscopic two-time quantities during
  aging.  (a) Auto-correlation function $C(t,t_w)$; (b)
  linear-susceptibility $\chi(t,t_w)$. The waiting time $t_w$
  increases from the left to the right curves, typically 
in a logarithmic scale such that, say, ${t_w}_{k+1}= 10 {t_w}_k$. The
  $\alpha$ relaxation time, $t_\alpha$, or the time-difference needed
  to decorrelate significantly, is an increasing function of $t_w$ and
  is indicated with an arrow in the left panel. The plateau in $\chi$
  occurs at $\chi _{\sc ea}=(1-q_{\sc ea})/T$ where $q_{\sc ea}$ is
  the height of the plateau in $C$.}
\label{fig-aging}
\end{figure}

Following this procedure one can study, for instance, the energy
relaxation as a function of $t_w$. However, such a macroscopic
`one-time quantity' does not capture all the richness of the
relaxation. Much more detailed information is contained in two (or
more) time-dependent global quantities.  In Fig.~\ref{fig-aging} we show the typical
behaviour of two {\it macroscopic} two-time quantities in many glassy
systems: the spontaneous de-correlation of a chosen global observable measured
at two subsequent times $t$ and $t_w$ (a), and the linear response of
the same observable measured at $t$ to a perturbation that
couples linearly to it between $t_w$ and $t$ (b).

As shown in panel~(a), the auto-correlation function exhibits a
two-step decay. At short time scales, say $t-t_w \ll t_w$, there
is a relaxation towards a plateau that we call the Edwards-Anderson
(EA) order-parameter $q_{\sc ea}$ (see its precise definition in 
Eq.~(\ref{eq-qea})).  The relaxation in this stage is
essentially independent of the waiting time $t_w$ and
time-translational invariance (TTI) holds $C(t,t_w)=C_{st}(t-t_w)$ to
a certain accuracy. The second relaxation at longer time scales, say $t-t_w
\gg t_w$, exhibits $t_w$-dependence and is not stationary. The latter
reflects the strong out-of equilibrium nature of 
aging.  The separation of time-scales is confirmed in panel~(b) where
we sketch the behaviour of the susceptibility $\chi(t,t_w)$.
The plateau occurs here at $\chi_{\sc ea}=(1-q_{\sc ea})/T$ with 
$q_{\sc ea}$ the value of $C$ at its plateau and $T$ the temperature of the 
environment that is henceforth 
measured in units of $k_B$, the Boltzmann constant. 

The relaxation of the correlation in Fig.~\ref{fig-aging}~(a) reminds
us, in a sense, of the two step relaxation in the supercooled liquid
phase -- the so-called $\alpha$ and $\beta$ relaxations~\shortcite{Donth}. Even more so,
it is basically indistinguishable from the one found in phase ordering
processes after a temperature quench (see Sect.~\ref{sec:coarsening}
for a detailed description of coarsening)~\shortcite{Bray94}.  In such processes domains of the
different equilibrium phases progressively grow in competition. At
each instant, the equilibrium order parameter is essentially uniform
within each domain.  When only short time-differences are explored the
domain-walls are basically static and the correlations decay just as
in equilibrium. Instead, when longer time-differences are reached the
walls move appreciably and the correlations
decay from the plateau in a manner that depends explicitly on the
waiting-time.  A key quantity in the description of
coarsening is the {\it typical domain linear size}, $L(t)$, which plays the
role of a dictionary between time $t$ and length $L$. All
properties of coarsening systems are invariant under rescaling of all 
lengths by $L(t)$ -- in a statistical sense.

\begin{figure}[h]
\vspace{0.5cm}
\centerline{
\includegraphics[width=0.9\textwidth]{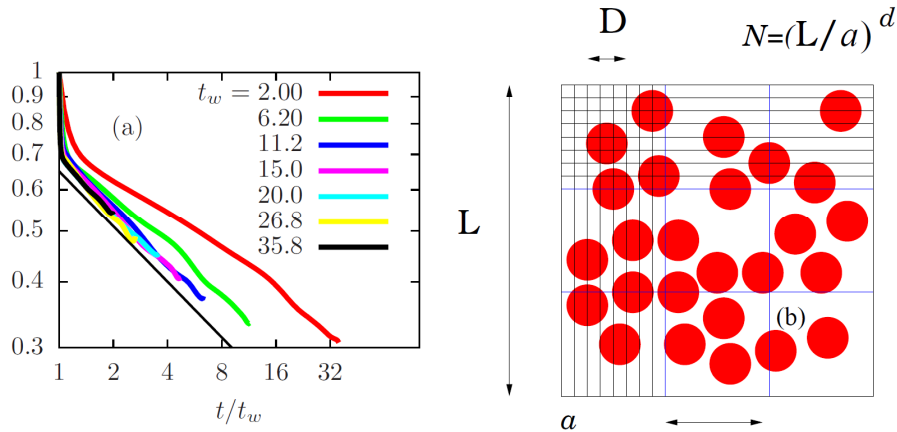}
}
\caption{(a) Scaling of two-time correlation data in an aging
  colloidal suspensions. Waiting times are given in the key in units
  of $10^2 $s.  The relatively good scaling with $t/t_w$ suggests
  $L(t) \simeq t^a$.  The data are taken from and analyzed along the
  lines explained in~\protect\shortcite{PNAS}. (b) A sketch of the pixelization
  used to describe the particle configurations with binary variables,
  $\sigma_\alpha=0,1$; and the coarse-graining procedure (see
    Sec.~\ref{spinlanguage}).  $a$ is the lattice spacing, $D$ the
  particle diameter, $\ell$ the coarse-graining length and $L$ the
  system size.}
\label{fig-aging-exp}
\end{figure}

The question naturally arises as to whether the dynamics in glassy
systems is also of coarsening type albeit the growing order had not
been identified yet. From the analysis of the correlations in many
glasses, and one example is shown in Fig.~\ref{fig-aging-exp} where
scaled data from an aging colloidal suspension are displayed~\shortcite{PNAS}, one
could easily conclude this to be the case. 
The figure shows
the two-time correlation 
data taken using different $t_w$s as a function of the ratio
$t/t_w$. For sufficiently long $t_w$ the scaling is quite good
suggesting that there might be a growing length $L(t)\propto t^a$,
with $a$ an undetermined power.  However, the scaling of the global
two-time correlation as $f(L(t)/L(t_w))$ is just a quite generic
property of two-time correlations~\shortcite{Cuku2}, and the space-time
correlations do not signal the growth of structure contrary to what happens in
simple phase ordering kinetics in which it scales as a function of
$r/L(t)$. Therefore, to reach a conclusion on the existence (or not)
of a growing length one should, ideally, analyze the dynamics at {\it
  mesoscopic time and length scales} or, at least, define and
explore more complex quantities giving access to the mesoscopic
details of the process.  Eventually, these studies might also
enlighten us about the reason for the growth.

Taking the case of usual phase ordering processes as a reference, we
realize that in order to settle this issue we need to follow a number
of basic steps summarized by the following questions: 1) how can we
define an order parameter for a glass? 2) how can we define domains 
-- within which the order parameter is essentially constant -- or
dynamically correlated volumes? 3) can one identify a growing linear
size and does it reflect the nature of the dynamics - {\it e.g.}
smooth (such as, for instance, the curvature driven kinetics observed
in some coarsening systems) vs. activated? 
4) can one understand the mechanism leading to such a growth?
In the body of this chapter we develop these questions. 

Having explained that the aim of this chapter is to discuss searches of 
growing lengths in {\it aging glasses}, we announce that 
we shall also deal with {\it a priori} simpler cases 
such as a particle's diffusion in random media or the motion of
a finite dimensional manifold, both in their aging regime. Although the glassy nature of these
problems might not be complete, it turns out that some aspects 
of their dynamics are similar to the ones encountered in
conventional glasses. Moreover, their mere definition is clearer
and
their analytic treatment as well as the identification of a growing length 
are simpler. Their discussion should then be instructive.

The organization of the chapter is the following. In
Sect.~\ref{sec:definitions} we give a number of definitions and 
we discuss how to deal with quenched disorder, if present.  In
Sect.~\ref{sec:coarsening} we review sub-critical and critical
coarsening phenomena in clean systems.  In this
Section we also introduce, briefly, a relatively
simpler problem with an easy to identify growing length: the motion of
a free elastic manifold.  Section~\ref{sec:droplet} is devoted to the
phenomenological decription of activated processes 
-- that are specially relevant in problems with quenched disorder -- 
starting with a
short account of aging in the Sinai model of difussion and
developing afterwards the droplet model predictions for the dynamics
of elastic manifolds in random media and spin-glasses.
Section~\ref{sec:glasses} recaps mean-field predictions for the
growing length issued from the study of fully-connected spin models
and the related mode-coupling theories as well as growth phenomena in
kinetically constrained models for glasses. We also discuss growing
lengths in other systems with aging dynamics: the interplay between
flow and internal relaxation and the effect of the former on the aging
properties of coarsening and glassy systems, granular matter and
quantum glasses.  Finally, in Sect.~\ref{sec:reparametrization} we
briefly recall a possible mechanism for dynamic fluctuations and the 
associated growing correlation length in aging
systems: the asymptotic development of time-reparametrization
invariance, that has been extensively reviewed elsewhere~\shortcite{chcu}.

We do not attempt to present a comprehensive review of analytic, numeric
and experimental studies of all kinds of aging systems. We rather focus 
on the search for  growing length-scales and their possible use to describe 
aging materials. With this in mind, the reader should not expect to find a complete
list of references on aging studies.
 
\section{Definitions}
\label{sec:definitions}

In this Section we summarize a number of definitions relevant to the
study of cooperative motion in glassy systems. These definitions do
not assume equilibrium and can be used to study out of equilibrium
and, in particular, aging
samples. They depend, in general, on various times independently and
no fluctuation-dissipation theorem is assumed. Although we are not 
going to dwell into the question of the pertinence of effective
temperatures in the description of glasses, we recall  its 
definition since it will appear in the rest of this Chapter.

\subsection{Euler and Lagrange descriptions}
 \label{spinlanguage}

Two ways of studying the heterogeneous dynamics in a many particle  
 (or many higher-dimensional object) system are the 
following.

\subsubsection{Lagrangian description}
 
One can follow the evolution of each individual particle, labeled by an index $i$,
and detect which are the fast and slow moving ones during a previously chosen time window,
say $t-t_w$, around some time, say $t_w$, after preparation. This is the route followed 
in early studies of dynamic heterogeneities in super-cooled liquids, see~\shortcite{Glotzer} 
and~\shortcite{Weitz} for a few of many molecular dynamics and confocal microscopy 
papers that use this type of measurement. One of the main outcomes of these 
studies is the observation of clustering of fast particles, in the form of strings
the length of which increases -- and may diverge -- close to the glassy arrest. 
In the trully glassy phase fewer studies exist. A precursor is~\shortcite{Miyagawa}.
More recently, Vollmayr-Lee {\it et al}
studied the geometry and statistical properties of clusters of mobile and inmobile
particles  with molecular dynamics~\shortcite{Vollmayr-Lee}. Confocal microscopy
has been used with similar aims~\shortcite{Courtland}.

\subsubsection{ Eulerian description}

In a first step, one can define Ising spin variables as the ones that describe 
uniaxial magnetic systems. The choice is done
for notation convenience but also because, as proposed in~\shortcite{PNAS},
a simple mapping between  particle positions and Ising spin
variables, $\sigma_\alpha$, defined on the $N=({\rm L}/a)^3$ vertices of a cubic
lattice captures the dynamics of `atomic' glassy systems as well (see 
Fig.~\ref{fig-aging-exp}-(b)). (${\rm L}$ is the linear length of the box and $a$ the 
length of the lattice spacing.) In a few words, one partitions space with a
lattice with very fine mesh -- smaller than the particle radii -- 
and the mapping assigns a spin one to each cell occupied by a piece of 
particle and zero otherwise. Once this is done, all observables are 
written in terms of the bimodal variables, as in spin models, and the identities of the particles
are ignored. 

In a second step, one can coarse-grain the spin variables over
boxes of a chosen linear size, $\ell$, with 
ideally ${\rm L} \gg \ell\gg a$.
The locality is then given by the position 
of the box which is labeled by an index $i$. The coarse-graining amounts to a partial 
averaging, the effective spin on each box, $s_i$, 
is now a continuous variable, 
$0 \leq s_i \leq 1$.

In a system with quenched randomness, realized
as fixed obstacles, local fields or else, 
the coarse-graining procedure  over a sufficiently large volume
averages over the peculiar local features of 
the fixed disorder. 
In this way,  the `finger-print'~\shortcite{castillo}
of disorder, that is to say local fluctuations 
determined by the particular local disorder -- such as Griffiths singularities in
random ferromagnets -- 
should be washed away. One  can then expect 
to arrive at a description of the noise-induced fluctuations present 
in problems with or without quenched disorder.

This method appears to be more adequate 
for analytic treatment through a field theory. 

\subsection{Space-time correlation} 
\label{space-time}

In usual coarsening systems, see Sect.~\ref{sec:coarsening}, 
the averaged space-time correlation function 
\begin{equation}
N C(r,t) =\sum_{ij/|\vec r_i-\vec r_j|=r}\langle \delta s_i(t) \delta s_j(t)\rangle
\label{facstrut}
\; , 
\end{equation}
with $\delta s_i(t) = s_i(t) - \langle s_i(t) \rangle
$ and $\langle s_i(t)\rangle=\langle s_i\rangle_{eq}$ in all the cases
we shall deal with, 
allows for the identification of a growing length from, for example,
 $L_{\rm a}(t) \equiv \int d^d r \ r^{{\rm a}+1} C(r,t) /$
\linebreak $\int d^d r \ r^{\rm a}
C(r,t)$.  (${\rm a}$ is a parameter chosen to weight preferentially short or long
distances; the time-dependence of $L_{\rm a}(t)$ should not depend on ${\rm a}$.)
Here and in the following $\langle \ldots \rangle$ stands
for an average over different realizations of thermal histories at
heat-bath temperature $T$ and/or initial conditions. In presence of
quenched disorder one adds an average over it and denotes it $[\dots
]$.  The stochastic time-dependent function $N^{-1}
 \sum_{ij/|\vec r_i-\vec r_j|=r}s_i(t) s_j(t)$ after a quench from
 a random initial condition does not fluctuate 
in the thermodynamic limit. Therefore, the averages are not really necessary 
but they are usually written down.
In spin-glasses and glasses this observable does not yield
information on the existence of any growing length as we shall discuss 
below.

\subsection{Two-time quantities} \label{two-time}

The auto-correlation function and linear susceptibility 
are defined as 
\begin{eqnarray} 
\begin{array}{rclcl}
N C(t,t_w)  &\equiv& \sum_{i=1}^{N}
\langle C_{i}(t,t_w) \rangle &=& \sum_{i=1}^{N} \langle \delta s_{i}(t) \delta s_{i}(t_w)
\rangle
\; ,
\label{eq-global-C}
\\
N \chi(t,t_w) &\equiv&\sum_{i=1}^{N}
\langle \chi_{i}(t,t_w) \rangle &=&\sum_{i=1}^{N}
\int_{t_w}^t dt' \; 
R(t,t')
\; , 
\end{array}
\end{eqnarray}
with $R_i(t,t') = \left. 
\delta \langle  s_i(t)\rangle/\delta h_i(t')\right|_{h=0}$
and $NR=\sum_{i=1}^N R_i$
the local and global instantaneous linear responses, respectively.

A number of fully general relations between the linear response and 
a correlation computed as an average over unperturbed system 
trajectories have been recently derived. Quite generally they read
\begin{eqnarray}
R(t,t_w) = 
\frac{1}{2T} \left[ \ \frac{\partial C(t,t_w)}{\partial t_w} \ 
+ N^{-1} \sum_{i=1}^N \langle s_i(t) B_i(t_w)\rangle
\right] \theta(t-t_w)
\; .
\label{eq:general-FDR}
\end{eqnarray}
The explicit form of the factor $B_i$ in the second term in the 
right-hand-side (r.h.s.) depends on the microscopic 
dynamics; it has been computed for Langevin processes~\shortcite{Cukupa} and Markov processes
for discrete variables~\shortcite{Lippiello05}. In all cases
$B_i$ is the deterministic drift in the sense that 
$\langle \dot s_i \rangle = \langle B_i\rangle$.

In equilibrium, the second term in the r.h.s. of  
Eq. (\ref{eq:general-FDR}) is equal to the first one, and
 $R$ (or $\chi$) and $C$ are related by the 
fluctuation-dissipation theorem
\begin{eqnarray}
&& 
R(t,t_w) = \frac{1}{T} \ 
\frac{\partial C(t,t_w)}{\partial t_w} \ 
\theta(t-t_w)
\; , \;\;\;
\\
&&
\chi(t,t_w) = \frac{1}{T} \ [C(t,t) - C(t,t_w)] \theta(t-t_w)
\; , 
\label{eq:linear-FDT}
\end{eqnarray}
and all two-time quantities depend on $t-t_w$ only. 

\subsection{Order parameter}

The very concept of a glass order parameter is not obvious. Although
some kind of static amorphous order may develop with time in a glassy
regime, searches have given negative results so far (not surprisingly
since one does not really know what is looking for)~\shortcite{Jorge}. The
simplest possibility, $\langle s_i(t)\rangle$, is void of information
and the space-time spin-spin correlation $C(r,t)$
%\left. \langle s_i(t) s_j(t)\rangle\right|_{|\vec r_i -\vec r_j|=r}$ 
(the Fourier 
transform of the structure factor) is, to a first 
approximation, time independent and very similar to the one in the 
super-cooled liquid~\shortcite{PNAS}.

\subsubsection{Edwards-Anderson order parameter}

A {\it dynamic} order parameter, named after Edwards and Anderson (EA)
who introduced it for spin-glasses, is defined as
\begin{equation}
q_{\sc ea}=\lim_{t \to \infty} \lim_{t_w \to \infty} C(t,t_w)
\; ,
\label{eq-qea}
\end{equation}
where $C(t,t_w)$ is the auto-correlation function.  The order of the
two long time limits is crucial since the weak long-term
memory~\shortcite{Bouchaud,Cuku1} ensures that $ \lim_{t_w \to \infty}
\lim_{t \to \infty} C(t,t_w)=0$ (see Fig.~\ref{fig-aging} (a)).  The
EA order parameter detects ergodicity breaking: in the liquid
(paramagnetic) phase $q_{\sc ea}=0$ because the system looses memory
at finite time scales but in the glass phase such memory remains. The
idea is quite generic and it can be applied to glassy
systems made of constituents of any kind.

\subsubsection{Replica overlap - a static counterpart}

The theory of spin-glasses~\shortcite{Mepavi} suggests the definition of a
{\it static} order parameter as the {\it overlap} between two replicas
(two systems with the same quenched randomness)
$a$ and $b$ subjected to a fictitious attractive coupling of strength
$\epsilon$ that forces them to be in the same thermodynamic state:
\beq 
q =
\lim_{\epsilon \to 0}\lim_{N \to \infty} 
\frac{1}{N}
\sum_{i=1}^{N} \langle s^{a}_{i} s^{b}_{i}\rangle_{\epsilon} \; .
\eeq 
The {\it spin-glass susceptibility} is its linear response to an
infinitesimal variation of the coupling: 
\beq 
\chi_{\sc sg} =
\left. \frac{\partial q}{\partial \epsilon} \right |_{\epsilon=0}=
\frac{\beta}{N} \sum_{i,j} \left( \langle s^{a}_{i} s^{a}_{j} s^{b}_{i}
s^{b}_{j}\rangle - \langle s^{a}_{i} s^{a}_{j} \rangle \langle
s^{b}_{i} s^{b}_{j}\rangle  \right)
\; .
\label{eq-chi-sg}
\eeq 
The averages $\langle \dots \rangle_\epsilon$ and 
$\langle \dots \rangle$
are taken here with the corresponding
Gibbs-Boltzmann distributions.
A crucial problem with
these definitions is that below $T_g$ aging persists at any
finite (with respect to some function of $N$) time scale.
Nonetheless, $\chi_{\sc sg}$ motivates the
definition of a dynamic analogue that could detect the growth of
static order during aging, see Sect.~\ref{subsec:static-growth}.

\subsection{Dynamically correlated volume}

A more natural proposal is to consider the spatial variation of the
{\it speed of relaxation} in different regions of space. This point of
view is very close in spirit to the idea of dynamical heterogeneity in
the supercooled liquid regime as characterized by local relaxation
times. As discussed later, this point of view also emerges naturally
in the theoretical analysis of soft-modes associated with the time
reparametrization invariance in the effective dynamical equations of
motion at long time scales~\shortcite{chcu}.   We present below different 
ways to access a correlation related to the local dynamics. 

\subsubsection{Spatial correlation of local dynamics} \label{spatial}

Let us regard the local
two-time dependent auto-correlation function $C_{i}(t,t_w)$ as the
{\it local order parameter}. Its spatial variation can be easily
characterized by the spatial correlation
function~\shortcite{castillo,parisi,jaubert} \beq
%G_{2,2}(i,j;t,t_w)
G_{4}(i,j;t,t_w)
=\langle C_{i}(t,t_w) C_{j}(t,t_w) \rangle - 
[C(t,t_w)]^{2}.
\label{eq-g4}
\eeq The subscript $4$ is to remind one that it is a 4-point function
since the auto-correlation function is already a two-point correlation
function (in time). 

The key question is whether there is a characteristic dynamical
length-scale, $\xi_{4}(t,t_w)$, beyond which $G_4$ 
de-correlates, {\it e.g.}
\beq 
-\ln G_{4}(i,j;t,t_w) \simeq
\frac{r}{\xi_{4}(t,t_w)}
+b \ln r
\label{eq:G4}
\eeq 
with $r=|\vec r_i - \vec r_j|$.
[This expression assumes the usual decay  $G_{4} \sim
r^{-b}\exp(-r/\xi_{4})$ but more
general forms of the type $G_4\simeq r^{-b}
f(r/\xi_4)$ have also been considered.]

By integrating Eq.~(\ref{eq:G4}) over space one obtains
\beq 
N
\chi_{4}(t,t_w)=
\sum_{i,j}
G_{4}(i,j;t,t_w) = 
\langle [ \sum_i C_i(t,t_w) ]^2 \rangle - 
\langle \sum_i C_i(t,t_w) \rangle^2
\label{eq-chi4}
\eeq 
which measures the dynamically correlated volume.
Note that in spite of being usually called a `susceptibility', strictly speaking
$\chi_4$ is not one~\shortcite{Semerjian,Lippiello08}. The same proviso applies to 
$\chi_{\sc sg}$ in Eq.~(\ref{eq:chiSG}). 

The 4-point correlation function in \eq{eq-g4} and the integrated one
in \eq{eq-chi4} are extensions of similar expressions defined for the
supercooled (stationary) liquid state. In the glassy regime one needs
to keep the $t_w$ dependence in $\xi_{4}(t,t_w)$. Still, it can be
measured in numerical simulations and in some experiments from
real-time real-space images.  One has to keep in mind, though, that
this quantity detects how similar motion in different regions is but
not necessarily whether these are ordering in the same state.

\subsubsection{Higher order susceptibilities}

Although $G_4$ has been studied in numerical
simulations~\shortcite{jaubert,JANUS} its direct experimental investigation
remains a challenge as, in general, multi-point correlators. (Lucky
exceptions are colloidal suspensions in which confocal microscopy
allows one to store the full particle configuration~\shortcite{PNAS}.)  A
natural way out would be to measure responses to external
perturbations, actual susceptibilities, as suggested in
\shortcite{Huse88,Bouchaud05} and done experimentally in \shortcite{Berthier05}.
The basic idea is that $G_4$ should be related to a non-linear
susceptibility by some sort of generalization of the FDT, much in the
same way as the ordinary correlation function is linked to the linear
susceptibility, see Eq.~(\ref{eq:linear-FDT}).  One could then measure
the latter to extract information on the former.  In order to fulfill
this program it is necessary to establish which are the non-linear
susceptibilities associated to multi-point correlators and which is
the generalization of the FDT, holding possibly out of
equilibrium.  Exact general relations between multi-point correlators
and non-linear response functions
\shortcite{Semerjian,Lippiello08,Bouchaud05} derived for systems subjected
to a Markovian dynamics show that beyond linear order the
susceptibilities are related not only to multi-spin correlations (such
as $G_4$) but also to more complicated correlators. For instance, 
for the second order response of two spins 
$R_{ij}^{(2)}(t,t_1,t_2)= \delta ^2\langle s _i(t)s _j(t)\rangle 
/\delta h_i(t_1)h_j(t_2)|_{h=0}$ in equilibrium one has 
\begin{eqnarray}
R_{ij}^{(2)}(t,t_1,t_2)&=&\frac{1}{2T} \left[ 
\frac{\partial}{\partial t_1}\frac{\partial}{\partial t_2}
\langle s _i(t)s_j(t)s_i(t_1)s_j(t_2)\rangle
\right . \nonumber \\
&& - \left . \frac{\partial}{\partial t_2}
\langle s_i(t)s_j(t)B _i(t_1)s_j(t_2)\rangle
\right ],
\end{eqnarray}
for $t_1\neq t_2$.  This feature poses the problem of choosing the
best suited non-linear susceptibility to detect cooperative
effects. In a series of papers both third~\shortcite{Bouchaud05} and
second~\shortcite{Lippiello08} order susceptibilities have been
considered. Analytical and numerical studies show that the non-linear
susceptibilities and $G_4$ obey analogous scaling forms from which one
can extract a cooperative length.  Experimental studies are on the
way~\shortcite{Caroline}.

The second order susceptibility $\chi ^{(2)}(t,t_w)=\int _{t_w}^t dt_1
\int _{t_w}^t dt_2 \ R_{ij}^{(2)}(t,t_1,t_2)$ is strictly related
to the fluctuations of $\chi_i
(t,t_w)$~\shortcite{Lippiello08,fluttuazioni}.

\subsubsection{Distributions of coarse-grained two-time quantities}

Another way of extracting a growing length, $\xi$, alternative to that
expressed by Eq.~(\ref{eq:G4}), is to study the full distribution of
local two-time
functions~\shortcite{castillo,Chamonetal1,Chamonetal2,Chamonetal3}.    
Indeed, for $\ell \gg \xi$ the
coarse-graining boxes naturally become independent and one should
recover a Gaussian distribution. The crossover from non-trivial to
trivial dependence can then be used to estimate $\xi$ that should,
presumably, behave as $\xi_4$. The
same argument can be applied to the pdf of the local linear
susceptibility $\chi_i$s.

\subsection{Growth of underlying static glass order?}
\label{subsec:static-growth}

An intriguing problem is whether any static glass order develops
during aging.  The spin-glass susceptibility, \eq{eq-chi-sg}, suggests
to define a 4-point correlation
function~\shortcite{JANUS,Huse88,rieger,rome,komori,yoshino}, 
\beq N \chi_{\sc
  sg}(t) = \sum_{i,j} G_{\sc sg}(i,j;t) = \sum_{i,j} \langle
s^{a}_{i}(t) s^{a}_{j}(t) s^{b}_{i}(t) s^{b}_{j}(t)\rangle \; .
\label{eq:chiSG}
\eeq 
Time $t$ is measured after the temperature quench at which the two replicas,
labeled $a$ and $b$,
are prepared in independent random initial configurations. One is
interested in finding a dynamic length-scale, $\xi_{\sc sg}(t)$,
beyond which $G_{\sc sg}$ decorrelates, 
\beq -\ln G_{\sc sg}(i,j;t) \sim
\frac{r}{\xi_{\sc sg}(t)} 
+c \ln r 
\; .  
\label{eq:xiSG}
\eeq 
 $\xi_{\sc sg}(t)$ is
simpler than $\xi_{4}(t,t_w)$ in that it is a one-time quantity like
the domain size $L(t)$ in usual phase ordering processes.

In the context of spin-glasses the value of the parameter $c$ 
is used to distinguish between a disguised ferromagnet picture, 
as proposed in the droplet model~\shortcite{Fisher-Huse}, and a more 
complex equilibrium structure, such as the one predicted
by mean-field models~\shortcite{Mepavi}. In the former $\lim_{r\to\infty} G_{\sc sg}(r)\to 
const$ and $c=0$ while in the latter $\lim_{r\to\infty} G_{\sc sg}(r)\to 0$ and 
$c>0$. 

It is interesting to compare $\chi_{\sc sg}(t)$ and
$\chi_{4}(t,t_w)$. Since there are no interactions between the two
replicas $a$ and $b$, purely dynamic correlation cannot exist
between them. Thus, the dynamical SG susceptibility $\chi_{\sc sg}(t)$
can only detect growth of (if any) static order much as domains in
usual phase ordering. On the other hand $\chi_{4}(t,t_w)$
can detect both static and dynamic order. 

\subsection{Effective temperature}

Although we shall not develop the $T_{\sc eff}$ ideas here we include a short
paragraph recalling its definition; the concept will appear in a number of 
places later in the article.

The deviation from the fluctuation-dissipation theorem (FDT)
found in mean-field glassy models~\shortcite{Cuku1,Cuku2} and in a
number of finite dimensional systems~\shortcite{Ritort} can be rationalized 
in terms of
the generation of an effective temperature~\shortcite{Cukupe} in the
system. The identification of a temperature out of equilibrium makes
sense in the asymptotic limit of small entropy production (either long
times or very weak applied drive, see Sec.~\ref{subsec:interplay}) 
in which the system evolves
slowly. $T_{\sc eff}$ is defined as minus the inverse slope of the asymptotic
parametric plot $\chi(C)$, where $\chi$ is the linear integrated
response and $C$ is the two-time correlation. The thermodynamic
character of $T_{\sc eff}$ has been checked in mean-field models and
mode-coupling theories as well as in simulations of Lennard-Jones
mixtures, models of silica and many others. A basic condition is that
$T_{\sc eff}$ obtained from different (interacting)
observables should be equal whenever
measured in the same dynamic regime.
A different scenario is found at the lower critical dimension and 
in critical dynamics, and in trap models with unbounded trap depths.

\section{Phase ordering}
\label{sec:coarsening}

Phase-ordering kinetics is an important problem for material science
but also for our generic understanding of pattern formation in
nonequilibrium systems.

Let us consider a physical macroscopic system in contact with an
external reservoir in equilibrium.  Imagine now that one changes a
parameter instantaneously in such a way that the system is taken from
a disordered phase to an ordered one in its (equilibrium) phase diagram.  
Two paradigmatic examples are
spinodal decomposition, {\it i.e.} the process whereby a mixture of
two or more substances separate into distinct regions with different
concentrations, and magnetic domain growth in ferromagnetic materials
quenched below the Curie temperature~\shortcite{Bray94}.

Closely related to the above is the process whereby a critical state
is approached {\it via} a quench from the disorder state right to the
temperature at which the phase transition occurs. 

In both sub-critical and critical coarsening the dynamical process starts
from an equilibrium high temperature disordered state and
progressively evolves building a new phase, stable at a lower
temperature, either ordered or critical.

\subsection{Growing length, aging and scaling}

The evolution of an initial condition that is not correlated with the
final equilibrium state (and with no bias fields) does not
reach equilibrium in finite times. More explicitly, domains of all
the phases of the equilibrium state at the final temperature $T$ keep
on growing, until their typical size $L(t)$ becomes of the order of
the system size ${\rm L}$.  For any shorter time the system is out of
equilibrium and, in particular, the non-equilibrium evolution does not
come to an end whenever the thermodynamic limit ${\rm L}\to \infty $ is
taken at the outset.

The very existence of a growing length $L(t)$ is at the heart of the
aging behavior observed in these systems as can be easily
understood. Since $L(t_w)$ increases with $t_w$ the system needs
larger rearrangements to decorrelate from older configurations.  This
simple fact is the origin of the strong $t_w$-dependence of the
autocorrelation and linear-response described in Sec.~\ref{sec:intro} and
depicted in Fig.~\ref{fig-aging}.

In the asymptotic time domain, when $L(t)$ has grown much larger than
any microscopic length in the system,  a {\it
  dynamic scaling} symmetry sets in, similarly to the usual scaling
symmetry observed in equilibrium critical phenomena.  According to
this hypothesis, the growth of $L(t)$ is the only relevant process and
the whole time-dependence enters only through
$L(t)$. Observables such as correlation and response functions take
precise scaling forms that will be discussed in
Secs.~\ref{growthcrit} and~\ref{growthsubcrit}.  Exceptional cases
where dynamic scaling is not observed will be shortly discussed in
Sec.~\ref{nonscal}.

\subsection{Aging at a critical point} \label{growthcrit}

The scaling behavior of binary systems quenched to the critical point
is quite well understood since this issue can be addressed via scaling
arguments \shortcite{Godreche02} and renormalization group approaches
\shortcite{Janssen89} which give explicit expressions for many of the
quantities of interest up to two loops order.  Numerical simulations
\shortcite{Lippiello06} confirm the analytic results and probe exponents
and scaling functions beyond the available perturbative orders.  In
this case the system builds correlated critical clusters with fractal
dimension $D=(d+2-\eta)/2$, where $\eta $ is the usual static critical
exponent, in regions growing algebraically as $L(t)\sim t ^{1/z_{eq}}$,
$z_{eq}$ being the dynamic equilibrium critical exponent relating times
and lengths.  

In the asymptotic time domain the correlation function (\ref{facstrut})
has the scaling form
\begin{equation}
C(r,t)=L(t)^{-2(d-D)}f\left( \frac{r}{L(t)}\right).
\label{facscalcrit}
\end{equation}
The pre-factor $L(t)^{-2(d-D)}$ takes into account that the growing
domains have a fractal nature (hence their {\it density} decreases
as their size grows) and the dependence on $r/L(t)$ in $f(x)$
expresses the similarity of configurations 
at different times once lengths are  measured
in units of $L(t)$. 

For two-time quantities, when $t_w$ is sufficiently large one has
\begin{equation}
C(t,t_w)=C_{st}(\tau) \ f_{\sc c}\left( \frac{L(t)}{L(t_w)}\right)
\label{cscalcrit}
\; ,\qquad\;
R(t,t_w)=R_{st}(\tau) \ f_{\sc r}\left( \frac{L(t)}{L(t_w)}\right).
\end{equation}
Here $C_{st}(\tau) \simeq L(\tau)^{-2(d-D)}$ and $R_{st}(\tau)
\simeq \beta L(\tau)^{-2(d-D)-z_{eq}}$, where $\tau =t-t_w$. 
The scaling functions $f_{\sc c}$ and $f_{\sc r}$ describe the
non-equilibrium behavior and take 
the limiting values $f_{\sc c}(0)=f_{\sc r}(0)=1$ and $f_{\sc c}(\infty) =
f_{\sc r}(\infty) =0$. The correlation
and response function of the equilibrium state at $T_c$ obey FDT,
$R_{st}(\tau)=\beta \dot C_{st}(\tau)$.
  In the scaling forms the
equilibrium and non-equilibrium contributions
enter in a {\it multiplicative} structure. Non-equilibrium
effects are taken into account by taking ratios between the sizes of
the correlated domains at the observation times $t_w$ and $t$ in the
scaling functions.  Of a certain interest is the limiting
fluctuation-dissipation ratio $T/T_{\sc eff}=X_\infty =\lim
_{t\to \infty} \lim _{t_w \to \infty} = TR(t,t_w)/[\partial
C(t,t_w)/\partial t_w]$, due to its universal character
\shortcite{Godreche02}.

Experiments on the non-equilibrium kinetics near a critical point are
reported in \shortcite{Joubaud09}, where the orientation fluctuations of
the director of a liquid cristal are measured after a sudden change of
the control parameter (in this case an AC voltage) from a value
in the ordered phase to one near the critical point
where the  Fr\'eedericksz second order
transition occurs. In this {\it quenching} procedure the
initial state is ordered.  Experimental data show a behavior of
two-time quantities in substantial agreement with the scaling pattern
described in Eqs.~(\ref{cscalcrit}), as
expected since the system can be described in terms of a
time-dependent Ginzburg-Landau equation, similarly to ordinary
magnetic systems. Interestingly enough, even the limiting
fluctuation-dissipation ratio $X_\infty$ turns out to be in good
agreement with the value found in the two-dimensional Ising model.

\subsection{Aging in the low-temperature phase}
\label{growthsubcrit}

The late stage of phase-ordering in binary systems is characterized by
a patchwork of large domains the interior of which is basically thermalized in
one of the two equilibrium phases while their boundaries are slowly
moving producing the power-law $L(t)\sim t ^{1/z}$. This
picture suggests the splitting of the degrees of freedom (spins) into
two categories, providing statistically independent contributions to observables such
as correlation or response functions. More precisely, a
quasi-equilibrium stationary contribution arises as due to bulk spins,
while boundaries account for the non-equilibrium part. 
Then~\shortcite{Bouchaud97}, asymptotically one has
\begin{equation}
C(r,t) \simeq C_{st}(r) + C_{ag}(r,t)
\label{eq:separation1}
\end{equation} 
The first term describes
the equilibrium fluctuations in the low temperature 
broken symmetry pure states  
\begin{equation}
C_{st}(r) = 
(1 - \langle s_i\rangle_{eq}^2) \ g\left ( \frac{r}{\xi _{eq}}\right ),
\end{equation}
where 
$\langle s_i\rangle_{eq}$ is the equilibrium expectation value of the 
local spin, and
$g(x)$ is a function with the limiting values $g(0)=1$, 
$\lim _{x\to \infty}g(x)=0$.
The second term takes into account
the motion of the domain walls through
\begin{equation}
C_{ag}(r,t) = 
\langle s_i\rangle_{eq}^2 \ f\left( \frac{r}{L(t)}\right),
\label{eq:C-aging1}
\end{equation}
with $f(1)=1$ and $\lim _{x\to \infty}f(x)=0$. 
Both
$C_{st}$ and $C_{ag}$ obey (separately) scaling forms with respect to
the equilibrium and the non-equilibrium lengths $\xi$, $L(t)$. 
In particular, Eq. (\ref{eq:C-aging1})
expresses the fact that system configurations at different times
are statistically similar provided that lengths are measured in
units of $L(t)$, namely the very essence of dynamical scaling.

An analogous {\it additive} separation holds for two time quantities
\begin{equation}
C(t,t_w) \simeq C_{st}(\tau) + C_{ag}(t,t_w),
\label{eq:separation}
\end{equation} 
and similarly for the response function.
The equilibrium character of the first term implies
\begin{equation}
C_{st}(\tau) = 
(1 - \langle s_i\rangle_{eq}^2) \ g_{\sc c}\left ( \frac{L(\tau)}{\xi _{eq}}\right ),
\end{equation}
where 
$g_{\sc c}(x)$ is a function with the limiting values $g_{\sc c}(0)=1$, $\lim _{x\to \infty}g_{\sc c}(x)=0$.
The non-equilibrium term obeys
\begin{equation}
C_{ag}(t,t_w) = 
\langle s_i\rangle_{eq}^2 \ f_{\sc c}\left( \frac{L(t)}{L(t_w)}\right),
\label{eq:C-aging}
\end{equation}
with $f_{\sc c}(1)=1$ and $\lim _{x\to \infty}f_{\sc c}(x)=0$. 
In the long $t_w$ limit
the two terms in Eq.~(\ref{eq:separation}) vary in completely
different two-time scales. The first one changes when the second one
is fixed to $q_{\sc ea} \equiv \langle s_i \rangle_{eq}^2$, see Eq.~(\ref{eq-qea}), at times such that
$L(t)/L(t_w) \simeq 1$. In this regime $C$ decays to 
a plateau at $q_{\sc ea}=\langle s_i\rangle^2$, see 
Fig.~\ref{fig-aging}. The second one varies when
the first one has already decayed to zero. The mere existence of the
second term is the essence of the aging phenomenon below $T_c$ with
older systems (longer $t_w$) having a slower relaxation than younger
ones (shorter $t_w$). Such a sharp separation of timescales is the
hallmark of quenches in the ordered phase, at variance with critical
quenches where equilibrium and non-equilibrium contributions are
entangled in the multiplicative form (\ref{cscalcrit}).  Although we
have used the terminology of binary systems, the scaling structure
discussed insofar is believed to hold quite generally
\shortcite{Bray94} (some exceptions will be discussed below), including
systems with more than two low temperature equilibrium phases (as
described, {\it e.g.} by Potts or clock models) or with a continuous
symmetry ({\it i.e.} vector $O(N)$ models).  Contrarily to the case of
critical quenches, however, a systematic expansion method in
sub-critical quenches is much more difficult~\shortcite{Mazenko04} and
general results comparable to those at $T_c$ are not yet available.
The scaling structure discussed above has been proven analytically
only in special cases, such as the one dimensional Ising chain with
Glauber dynamics~\shortcite{Godreche02,Lippiello00} or the Langevin dynamics
of the $d$-dimensional $O(N)$ model with non-conserved order parameter
in the large $N$ limit \shortcite{Corberi02}.  It is supported by
semi-analytical arguments in two dimensions~\shortcite{Sicilia,Sicilia-EPL},
numerical simulations~\shortcite{Simulscaling,Simulscaling2} and approximate 
theories~\shortcite{Berthier99}.  It has also been confirmed 
in experiments~\shortcite{Chou81}.

\subsubsection{The growing length}

The growing length depends on a few characteristics of the ordering
process -- the dimension of the order parameter, whether there are
conservation laws, the presence of quenched disorder -- and may serve
to classify systems in classes akin to universality
ones~\shortcite{Bray94}. Although the following results are hard to obtain
with rigurous arguments, they are by now well established.  Basically,
one distinguishes two important cases: in clean models $L(t) \simeq
t^{1/z}$; in problems with quenched randomness, where topological
defects are pinned, the growing length is expected to slow down from a
power law to a logarithmic dependence on time (due to thermal
activation above barriers with a power-law distribution) see
Sect.~\ref{sec:droplet}.  The dynamic exponent $z$ depends on the type
of microscopic dynamics and dimension of the order parameter. For
example, in curvature driven growth with scalar order parameter $z={1/2}$ and in 
phase separation with scalar order parameter (and no hydrodynamics) 
$z=1/3$. A crossover at a static length explains how the
activated scaling can be confused with a temperature-dependent power
law in dirty cases~\shortcite{Bustingorry} as explained in
Sect.~\ref{sec:droplet}.

\subsubsection{The scaling functions}

The full theoretical description of a coarsening process necessitates the 
determination of the scaling functions. This is a hard task and there is no
 powerful and systematic method to attack this problem  yet.

 Still, Fisher and Huse proposed that the scaling functions in the 
 aging regime, {\it e.g.} $f_{\sc c}$,  should be
 robust~\shortcite{Fisher-Huse}.  Changes in the model
 definition that do not modify the nature of the equilibrium initial
 nor target ordered state -- such as weak quenched disorder not
 leading to frustration -- should not alter the scaling functions. This is
 the so-called {\it super-universality hypothesis}. In this way the
 scaling functions in spin models with random ferromagnetic bonds or
 random fields should be identical to those found in the pure limit. Numerical 
tests in $d>1$ systems point in the direction of validating this 
hypothesis~\shortcite{Sicilia-EPL,Henkel-Pleimling,Simulscaling2} while 
very recent studies of the scaling properties of the linear response
in the $d=1$ random bond ferromagnet tend to falsify it~\cite{Puri}.

\subsubsection{The four-point correlation $G_4$ }

\begin{figure}
    \centering
    %\vbox to 8.5 cm {
   \rotatebox{0}{\resizebox{.45\textwidth}{!}{\includegraphics{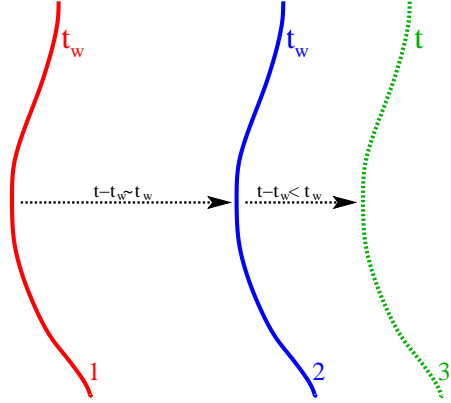}}}
   \vspace{0.5cm}
    \caption{Schematic representation of the mechanism of growth and
saturation of $\chi_4(t,t_w)$ in coarsening systems.}
\vspace{1cm}
\label{schemino}
\end{figure}

The behavior of $G_4$, and the way $L(t)$ is encoded in it, can be
easily understood in coarsening systems. In Fig.~\ref{schemino} a
configuration of the system with two interfaces (denoted 1 and 2,
continuous lines) at time $t_w$ is sketched.  The dashed interface
denoted as 3 is the location of interface 2 at the later time $t$
(such that $t-t_w<t_w$). For $t-t_w \ll t_w$ one has $C_i(t,t_w)=1$
-- for concreteness we think in terms of Ising spins --
everywhere except in the regions spanned by an interface in the
interval $(t_w,t)$ -- the region between 2 and 3 in
Fig. \ref{schemino} -- where $C_i(t,t_w)=-1$. Hence $G_4$ decays on a
volume of order $L^d(t)-L^d(t_w)$ which grows in time, and $\xi _4$
increases. For longer time differences ($t-t_w \stackrel{>}{\sim}
t_w$), however, another interface present at $t_w$ (1 in the sketch) may
superseed at time $t$ the position of interface 2 at the previous time
$t_w$. From this time on, $\chi _4$ stops growing and saturates to a
value of order $L^d(t_w)$, namely the typical volume contained between
two interfaces at $t_w$ ({\it i.e.} between 1 and 2).  The saturation can
also be explained by observing that for long time differences, $t-t_w \gg t_w$,
$\chi _4 $ factorizes as $(1/N)\sum _{i,j}\langle s_i(t)s_j(t)\rangle
\langle s_i(t_w)s_j(t_w)\rangle$.  Enforcing scaling, $\langle
s_i(t)s_j(t)\rangle\simeq f[r/L(t)]$, one recovers $\lim
_{t-t_w\to \infty} \chi_4(t,t_w)\propto L^d(t_w)$.  In this way,
$L(t_w)$ can be extracted from the long $t-t_w$ behavior of $\chi _4$
but, of course, this is not really necessary in such coarsening
problems. Importantly enough, $L(t)$ is a measure of growth of static
order in these cases.

\subsubsection{Exceptional cases: breakdown of dynamic scaling} \label{nonscal}

There are also a number of coarsening systems
 in which dynamic scaling breaks down
\shortcite{Bray94}.  Among these are the XY model [$O(2)$ symmetry] in
one~\shortcite{Bray95} and two~\shortcite{Bray00} dimensions, the clock model
with $p> 4$ states \shortcite{Andrenacci06}, the one-dimensional Heisenberg
model [$O(3)$ symmetry] \shortcite{Burioni09} and the large-$N$ model with conserved
dynamics \shortcite{Coniglio89}. Interestingly, in most cases this is
due to the presence of more than one length growing macroscopically
during phase ordering.  Let us consider the XY model in $d=1$ as a
simple paradigm. The order parameter in this case is a planar
vector. For long times after a quench from a high to zero 
temperature, when the excess energy is greatly reduced, the
continuous nature of the order parameter allows only soft (Goldstone)
modes, namely smooth rotations of the order parameter, called
textures. A winding length $L_w$ can be defined as the typical
distance over which a $2\pi$ rotation of the spin is detected. In
addition, due to the symmetry of the disordered initial state and of
the Hamiltonian between clockwise (textures) and counterclockwise
(anti-textures) rotations, both topological defects are formed
in the evolution, separated by winding inversions. Denoting with $L$
the typical distance between winding inversions, it is clear that
the 
reduction of energy implies the growth of both $L_w$ and $L$ as time
elapses. It was shown \shortcite{Bray95} that these lengths grow with
different exponents and this phenomenon prevents dynamic scaling to
set in.

\subsection{Elastic manifolds}
\label{sec:elasticline}

The dynamics of a directed $d$-dimensional elastic manifold embedded
in an $N$ dimensional transverse space is simply modelized by a
Gaussian scalar field theory that goes under the name of
Edwards-Wilkinson (EW) model. Non-linear effects can be accounted for by
including a Kardar-Parisi-Zhang (KPZ) term.  It has been
known for some time that these systems age 
and that the aging dynamics is governed by a
growing correlation length~\shortcite{Bray94,Cukupa,elastic-line2,KPZ}.

The aging regime, before saturation at a time-difference that depends
on the length of the line, ${\rm L}$, is characterized by a multiplicative scaling
of two-time quantities. For instance,  a two-time generalization of the roughness,
$\langle w^2\rangle (t,t_w)\equiv L^d \int d^d x 
\langle [\delta h(\vec x, t)-\delta h(\vec x,t_w)]^2\rangle$,
with $\delta h(\vec x, t) = h(\vec x, t) - \langle h(\vec x, t)\rangle$, 
$h$ the height of the manifold and $\vec x$ the position on the $d$-dimensional 
substrate, scales as 
\begin{eqnarray}
\langle w^2\rangle 
(t,t_w) \simeq L^{2\zeta}(t_w) \ f_{w^2}\left(\frac{L(t)}{L(t_w)} \right) 
\; . 
\end{eqnarray}
Assuming $\lim_{x\to\infty} f_{w^2}(x) = x^{2\zeta}$, in the limit
$\tau \gg t_w$ this observable reaches a stationary regime in which
$\langle w^2 \rangle (t, t_w) \simeq L^{2\zeta}(\tau)$.  For even
longer time-delays such that $L(t) \to {\rm L}$ one finds saturation
at ${\rm L}^{2\zeta}$.  The roughness exponent $\zeta$ is due to
thermal fluctuations and it is simply equal to $(2-d)/2$ in the EW
manifold.  The cross-over to saturation is then described by an
extension of the Family-Vicsek scaling that takes into account the out
of equilibrium relaxation and aging effects.  Correlation and response
functions are related by a modified fluc\-tuation-dissipation relation
after removing the diffusive factor and a well-defined effective
temperature~\shortcite{Cukupe} exists and depends on the initial
($T_0$) and final ($T$) values of the temperature before and after the
quench with $T_{\sc eff}>T$ if $T_0>T$ and $T_{\sc eff}<T$ if
$T_0<T$~\shortcite{elastic-line2,KPZ}.

\section{Role of activation: the droplet theory}
\label{sec:droplet}

The dynamics of glassy systems at low enough temperatures should
be dominated by thermal activation. Although it is very
difficult to study activated processes from first principles, several
phenomenological proposals for models with quenched disorder 
exist.  In particular, a droplet picture has been put forward for spin-glasses
\shortcite{Fisher-Huse,BM87,Fisher-Huse-static} and related systems
including elastic manifolds in random media \shortcite{Fisher-Huse-DPRM} and
vortex glasses \shortcite{vortexglass}. This model assumes that a static low
temperature phase, associated with a zero temperature glassy fixed
point in a renormalization group sense, exists in these
systems. Droplet-like low-energy excitations of various sizes $L$ on
top of the ground state render the dynamics strongly heterogeneous
both in space and time. The typical free-energy gap of a droplet with
respect to the ground state and the free-energy barrier to nucleate a
droplet are assumed to scale as $L^{\theta}$ and $L^{\psi}$,
respectively, with $\theta$ and $\psi$ two non-trivial exponents.
Static order is assumed to grow as in standard coarsening
systems. Dynamical observables such as the two-time auto-correlation
function should then  follow universal scaling laws in terms of a
growing length $L(t)$ originated in Arrhenius activation over 
barriers growing as a power of the length 
\begin{eqnarray}
t \sim \tau_0 \ e^{\frac{L^\psi}{T}} \qquad \Rightarrow \qquad 
 L(t) \simeq [T \ln (t/\tau_0)]^{1/\psi}
\; . 
\label{eq:Arrhenius}
\end{eqnarray}
The strong-disorder renormalization approach in configuration space~\shortcite{Monthus}
yields an explicit construction in favor of the droplet logarithmic scaling.

In the case of spin-glasses the very existence of static glassy states
is accepted but their detailed nature is a much debated
issue. Moreover, it is far from obvious whether coarsening of 
only two competing states occurs and
whether the growth is determined by thermal activation over such
barriers. In the present section we discuss a recipe to examine the
droplet picture quantitatively.

\subsection{Efficient strategy for data analysis}

In practice, the asymptotic dynamic scaling features associated with
the putative zero temperature glassy fixed point are difficult to
access in numerical simulations and experiments.  The following
strategy helps avoiding the control of pre-asymptotic effects that
last for very long~\shortcite{HY-unpublished}:
\begin{itemize}
\item[I)] Measure the dynamical length $L(t)$ and analyze it
over the full time duration
by taking care of the crossover from the initial
non-activated dynamics, that could be diffusive, critical or else, to
the asymptotic activated regime.
\item[II)] Reparametrize the time-dependent quantity of interest, say
$A(t)$, using $L(t)$ obtained in I) as a time-length dictionary. 
\end{itemize}

In this way pre-asymptotic corrections are dealt with separately:
those due to $L$ and those due to the scaling functions.  Once the
`dictionary' $L(t)$ is determined, step-II) is very much
straightforward: no uncontrolled fits are needed since the essential
exponents (such as the energy exponent $\theta$) are provided by
independent studies of static properties.  We prove the
efficiency of this strategy by analysing numerical results for Sinai
diffusion~\shortcite{sinai} and we recall the study of the random manifold
and Edwards-Anderson spin-glass along these lines.

\subsubsection{ Sinai model:  a test case}

The Sinai model is a random walker hopping on a one-dimensional lattice
$i=1,2,\ldots N$ over which a quenched random potential $U_{i}$ is
defined. The statistics of the random potential are such that
$[(U_{i}-U_{j})^{2}]=r_{ij}$ where $r_{ij}$ is the distance 
between sites $i$ and $j$ and $[\cdots]$ stands
for the average over different realizations of the random
potential. The statistical analysis of disorder yields
$\theta=\psi=1/2$.

The walker starts from a randomly chosen initial point at time
$t=0$. The mean-squared displacement is ${\rm B}(t,t_w)=[\langle
(x(t)-x(t_w))^{2} \rangle]$ with $x(t)$ the position at time $t$ and
$\langle \cdots \rangle$ an average over thermal histories and initial
conditions. The rigorous analysis of $\sqrt{{\rm B}(t,0)}=L(t)$ gives the
diffusion law $L(t) \propto [T \ln(t/\tau_{0})]^{2}$ at temperature
$T$~\shortcite{sinai}. Also of interest is the linear-susceptibility,
$\chi(t,t_w)$, of the averaged position with respect to a small bias
field acting on the walker after the waiting time $t_w$. In equilibrium
the FDT relates the two quantities as $\chi(t,t_w)={\rm B}(t,t_w)/2T$.

The asymptotic scaling properties have been almost fully uncovered by
a real-space renormalization group (RSRG) approach \shortcite{sinai-rsrg}
which is believed to become exact asymptotically. For illustrative
purposes, we reproduce some of these results with
droplet scaling arguments.

In the quasi-equilibrium regime $L(\tau=t-t_w) < L(t_w)$
the RSRG predicts ${\rm B}(\tau+t_w,t_w) = 2 T \chi(\tau+t_w,t_w) \sim T L^{3/2}(\tau) g_{\sc b}(L(\tau)/L(t_w))$ 
where $g_{\sc b}$ is a
scaling function. The exponent $3/2$ can be explained as follow. In
first approximation at time $t_w$ the particle sits in the lowest
energy minimum within a length scale $L(t_w)$ around the initial
point. Since $L(\tau) \ll L(t_w)$ the particle does not have time to
explore regions which are far from this minimum and diffusion
typically vanishes in the quasi-equilibrium regime. However, with a
small probability $\delta U_H/L^\theta(\tau) \sim T/L^{\theta}(\tau)$
with $\theta=1/2$ there is a secondary energy minimum within the
length scale $L(\tau)$ with the energy barrier between them being lower
than the thermal energy $T$.  In such a rare event the particle can hop
to the secondary minimum yielding the disorder-averaged behaviour ${\rm B}(\tau) \simeq T/L^{1/2}(\tau) L^2(\tau) \simeq T L^{3/2}(\tau)$.

In the aging regime $L(t) > L(t_w)$ the particle diffuses over longer
and longer lengths looking for lower and lower energy minima. Within a
given time scale $t$ the particle moves to the right or
to the left  by an amount $L(t)$. However, the direction of motion is 
almost deterministically given by the direction with the lowest
energy barrier since the difference between energy barriers (not minima)
is of order $L^{1/2}(t)$, a diverging quantity in the long times limit.
One then has ${\rm B}(t,t_w) \sim L^2(t)f_{\sc b}(L(t)/L(t_w))$.
This also implies that the linear susceptibility
typically vanishes in the aging regime since a change of the
potential of order $\delta U_H=h L(t)$ (an analogue of the Zeeman energy in
magnets) induced by an {\it infinitesimal} field $h$ does not affect
the difference in energy barriers of order $L^{1/2}(t)$. 
However, there are rare events
such that the energy barriers to the left and right are almost
degenerate with only a small difference in height 
$\delta U_H$. The probability to find
such degenerate barriers scales as $\delta U_H/L^{\psi}$ with
$\psi=1/2$. The infinitesimal $h$ then induces a change in 
the direction of diffusion resulting in a disorder-averaged displacement 
of order $(\delta U_H/L^{\psi}) L$ and thus
 $\chi(t,t_w) \sim L^{3/2}(t)f_{\chi}(L(t)/L(t_w))$.

\begin{figure}[h]
\vspace{0.5cm}
\begin{center}
\includegraphics[width=0.49\textwidth]{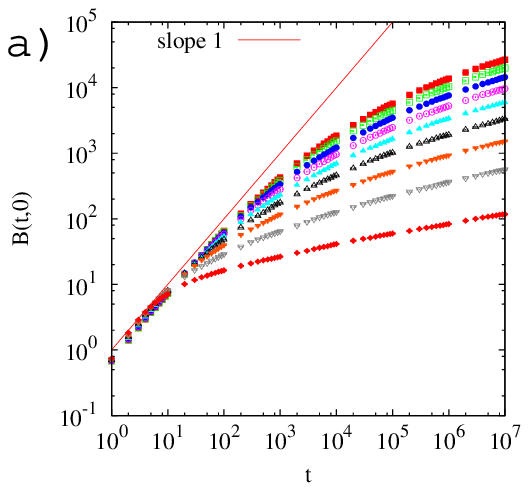}
\includegraphics[width=0.49\textwidth]{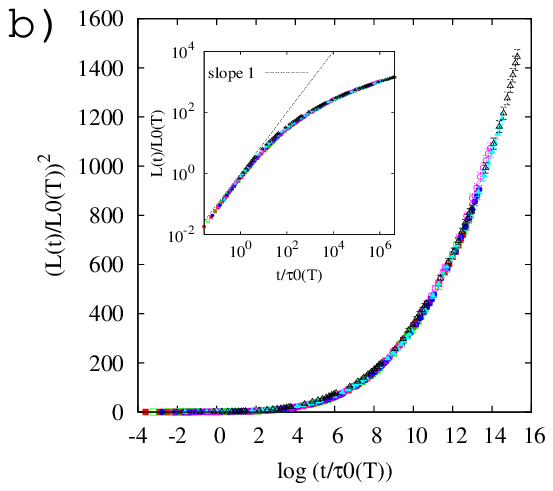}
\end{center}
\caption{Diffusion in the Sinai model: 
the mean-squared displacement from MC simulations.  a) Raw data
at $T=1.8,1.6,\ldots,0.4,0.2$ from top to
bottom. The unit of time $t$ is one MCs. The average is taken over
$10^{5}$ realizations of the random potential.  b) Scaling plot of the
dynamic length $L^2(t)={\rm B}(t,0)$ showing the
crossover behaviour from normal diffusion at short time
$t$ -- see the inset -- to the expected $L^2 \simeq \ln^4(t/\tau_0)$
asymptotic law.}
\label{fig-sinai-diffusion}
\end{figure}

\begin{figure}[h]
\begin{center}
\includegraphics[width=0.46\textwidth]{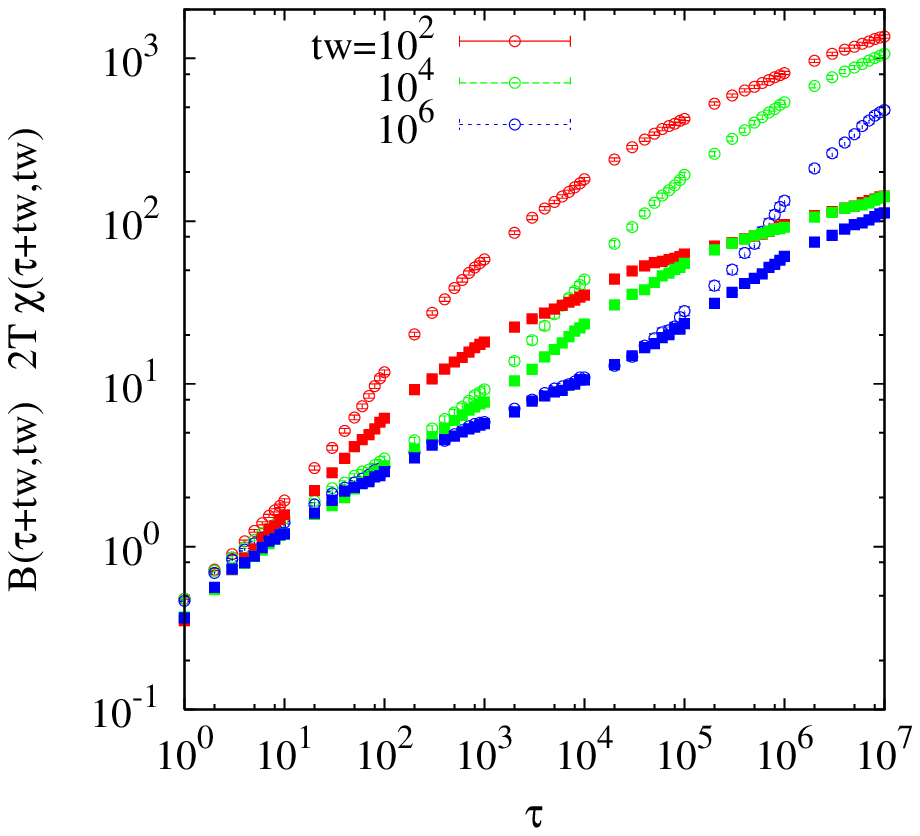}
\includegraphics[width=0.51\textwidth]{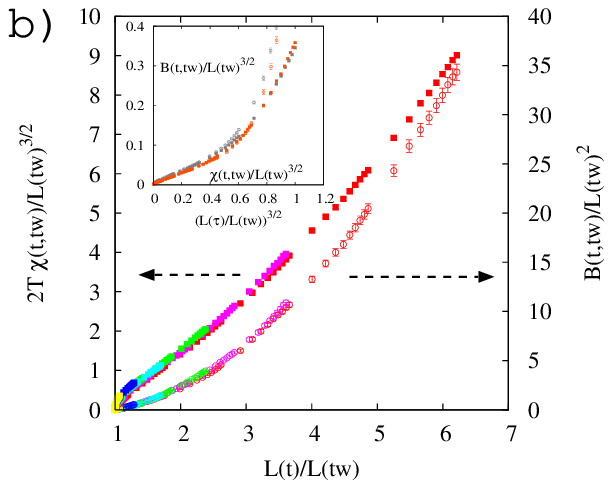}
\end{center}
\caption{Diffusion in the Sinai model: 
two-time mean-squared displacement
and linear susceptibility from MC simulations at $T=0.6$.
Note the absence of a developing plateau. 
a) Raw data of the
mean-squared displacement ${\rm B}(t,t_w)$ (open symbols) and the
linear-susceptibility $\chi(t,t_w)$ (filled symbols). 
b) Scaling plots  in
terms of the dynamic length. Inset: zoom over 
the quasi-equilibrium regime.}
\label{fig-sinai-aging}
\end{figure}

We performed Monte Carlo (MC) simulations to examine the anticipated
asymptotic scaling~\shortcite{HY-unpublished,deCandia}. In
Fig.~\ref{fig-sinai-diffusion} a) we show the mean-squared
displacement ${\rm B}(t,0)$ of the Sinai walker at different temperatures,
decreasing from top to bottom. It defines unambiguously a dynamical
length $L(t)=\sqrt{{\rm B}(t,0)}$ that we shall use below to reparametrize
time dependent quantities.  The data demonstrate that extremely long
times are needed to reach the $L(t) \simeq \ln^2 t/\tau_0$ asymptotic result
known to be exact analytically.  At finite $T$ the short-time
diffusion is normal ${\rm B}(t,0)=Dt$ with a diffusion constant $D$, see the
inset in Fig.~\ref{fig-sinai-aging}.  The crossover to the activated
regime takes place at a `thermal length scale' $L_{0}(T)$ such that
$\delta U(L_{0}(T))=L^{1/\theta}_{0}(T) \sim T$ and the corresponding
time scale $\tau_{0}(T)$ is fixed using
$L^{2}_{0}(T)=D\tau_{0}(T)$. As shown in
Fig.~\ref{fig-sinai-diffusion} b) the slow crossover is well described
by proposing that the scaled length $L(t)/L_{0}(T)$ is a universal
function of the scaled time $t/\tau_{0}(T)$. The asymptotic Sinai's
diffusion law can be parametrized precisely with $L(t) \sim L_{0}(T) [
\ln(t/\tau_{0}(T)) ]^{2}$.

Next we proceed to step II). In Fig.~\ref{fig-sinai-aging} a) we
show the aging effects observed in the mean-squared displacement and
the corresponding linear-susceptibility. In 
Fig.~\ref{fig-sinai-aging} b), the numerical data are reparamerized by the
dynamical length scale obtained in the step I) and follow the
expected scaling laws. This means that the correction to the
asymptotic behaviour is not large in terms of length scales.
The absence of a plateau in $B$ and $\chi$ in their raw and 
scaled forms demonstrates the fact that there is no 
additive separation of time-scales -- as in Eq.~(\ref{eq:separation}) -- in this problem.

\subsubsection{Elastic manifolds in random media}

The dynamics of a directed $d$-dimensional elastic manifold embedded
in an $N$ dimensional transverse space under the effect of a quenched
random potential play an important role in a variety of physical
systems ranging from coarsening in dirty systems to fracture.  An
application is the one in which the directed lines are vortices in
super-conductors aligned in the direction of the magnetic field and
simultaneously pinned by impurities.

These systems are intimately related to the problem of Sinai's
diffusion. The simplest case is a 1+1 dimensional directed polymer in
random media (DPRM). Each configuration is described by the transverse
displacement $x(z,t)$ at position $z$ along the directed polymer at
time $t$. Two natural quantities used to characterize aging are the
mean-squared displacement ${\rm B}(z,t,t_w)=[\langle(x(z,t)-x(0,t_w))^{2}
\rangle]$ and the linear susceptibility $\chi(z,t,t_w)$.  The FDT
$\chi(z,t,t_w)={\rm B}(z,t,t_w)/2T$ is satisfied in equilibrium.

Scaling arguments of droplet type  naturally
apply to the present case \shortcite{HY-unpublished}.  To this end one just
needs to keep in mind the scaling relation between the longitudinal
length $L$ and transverse length $L_\perp \propto L^{\zeta}$
 with $\zeta$
the roughness exponent.  Depending on the explored length scale, the
latter takes a thermal, $\zeta_T=1/2$ or a disorder, $\zeta_D=2/3$, value
with the crossover given at a static temperature dependent length
$L_0(T)$.  The disorder-dominated roughness exponent is related to the
energy exponent $\theta$ by an exact scaling relation
$\theta=2\zeta_D-1$ implying $\theta=1/3$. It is also conjectured that $\psi=\theta$. 
The time evolution of a local segment $x(z,t)$ may be viewed as
diffusion of a particle which feels an effective (renormalized)
potential created by the rest of the system, the variation of which
scales with the transverse length as
$\Delta U \simeq L^\theta \simeq L_\perp^{\theta/\zeta_D} \simeq L^{2-1/\zeta_D}_\perp$ in the disorder
dominated regime~\shortcite{LeDoussal-Vinokur}.

Using the analogy with the Sinai model one derives the scaling behaviour of different
observables. 
At a time $t$ after the quench, the system is equilibrated up to
longitudinal length $L(t)$ over which the energy is higher
than the equlibrium one by an amount of order $L^{\theta}$.
Thus the energy density per unit longitudinal length, $e(t)\equiv U(t)/L$, is expected to 
decay as
$e(t)=e(\infty)+{\mbox ct}/L(t)^{1-\theta}$.
In the quasi-equilibrium regime
the consideration of the degeneracy of minima suggests
${\rm B}(z,\tau+t_w,t_w)=2T\chi(z,\tau+t_w,t_w)=T L(\tau)g_{\sc b}(z/L(t_w),L(\tau)/L(t_w))$ [the prefactor 
$TL(\tau)$ is due to $T/L^\theta(\tau) L^{2\zeta}(\tau) = TL(\tau)$ since $\theta=2\zeta-1$.]  
In the aging regime we find
${\rm B}(z,t,t_w)=L(t)^{2\zeta}f_{B}(z/L(t_w),L(t)/L(t_w))$ while
$\chi(z,t,t_w)=L(t)f_{\chi}(z/L(t_w),L(t)/L(t_w))$ due to the
degeneracy of barriers.
 
In order to test the above, the 
simplest protocol is to choose a flat initial condition,
{\it i.e.} $x(z,0)=0$ for all $z$. During isothermal aging the
roughness of the system develops progressively from short
 to long wave lengths. The dynamical length $L(t)$ can
be extracted from the growth of the static roughness
${\rm B}(0,t,t)=[\langle(x(z,t)-x(z,t)^{2} \rangle]=B_{\sc
eq}(z=L(t))$. Here $B_{\sc eq}(z)=\lim_{t \to \infty}{\rm B}(z,t,t)$ is the
equilibrium roughness, which can be computed numerically with a
transfer matrix method in the 1+1 case~\shortcite{HY-unpublished}.
Another way to determine $L(t)$ is given in~\shortcite{Bustingorry}.

By performing Monte Carlo simulations of the 1+1 DPRM the analysis of
step I) and II) can be done
precisely~\shortcite{HY-unpublished,Bustingorry} helped by the knowledge of
the exact values of the exponents $\zeta_T=1/2$, $\zeta_D=2/3$ and
$\theta=1/3$.  Quite interestingly the variation of the energy scales
as $\Delta U\simeq L_\perp^{1/2}$ just as in the Sinai model discussed
before.

Concomitantly with the change in roughness exponent from 
thermal to disorder dominated values, 
$L(t)$ exhibits a gradual crossover from pure diffusion with $L(t)
\simeq t^{1/z}$ (and $z=2$) to an activated regime consistent with an
algebraic growth of barriers~\shortcite{Fisher-Huse-DPRM}. This is similar
to what is found in the Sinai model (see
Fig.~\ref{fig-sinai-diffusion}) but with $\psi=1/3$. The crossover
occurs at a static temperature-dependent correlation length $L_0(T)$. In
the analysis of numerical data the two regimes tend to be confused
into a single one with an effective temperature-dependent power-law,
$L(t) \sim t^{1/\overline z(T)}$~\shortcite{elastic-line}, the
$T$-dependence of which is inherited from the one in
$L_0(T)$~\shortcite{Bustingorry}, but this is just an approximation. 
Further support to the asymptotic logarithmic scaling is
given by the renormalization group study in~\shortcite{Monthus}.

 As regards the scaling properties of the two-time quantities
${\rm B}(z,t,t_w)$ and $\chi(z,t,t_w)$ in the quasi-equilibrium and aging
regimes one also faces the difficulty of going beyond the thermal
regime and reaching, for sufficiently long time-scales, the disorder
dominated one. In~\shortcite{Bustingorry} it was shown that in
the early effective power-law regime, approximated by $L(t) \simeq t^{1/\overline
  z(T)}$, and using the thermal roughness exponent, two-time
linear response and correlation functions conform to the scalings
discussed above and a finite effective temperature exists [basically
because $2\zeta_T=1$ and the diffusive prefactors in $B$ and $\chi$
are both equal to $L(t)$]. Simulations entering the trully activated
regime suggest that the dynamic scaling above with the disorder
roughness exponent $\zeta_D$ sets in (and the effective temperature
progressively vanishes)~\shortcite{HY-unpublished}.

\subsubsection{ Spin glasses}

The isothermal aging of the Edwards-Anderson (EA) model was analyzed along
the same steps.  The model is supposed to have a finite temperature phase 
transition and thus an additive separation of time-scales of the 
form~(\ref{eq:separation}) -- the precise nature of the aging term 
depending on a two-state droplet picture~\shortcite{Fisher-Huse,BM87,Fisher-Huse-static} or a more complicated 
dynamics of the Sherrington-Kirkpatrick (SK) type~\shortcite{Cuku2}.  

The growing length is extracted from the real-replica
overlap, see eq.~(\ref{eq:xiSG})~\shortcite{Huse88}.  It is common to use the  fit  $\xi_{\sc sg}(t)\sim
t^{1/\overline z(T)}$ with $\overline z(T)$ an effective temperature dependent
exponent~\shortcite{JANUS,Huse88,rieger,rome,komori}.  The best currently
available data are from the JANUS collaboration suggesting $\overline
z(T)\simeq\overline z(T_c) T_c/T$~\shortcite{JANUS}.  However, $\xi_{\sc sg}(t)$
should 
exhibit a gradual crossover from critical dynamics at short-times to
activated dynamics asymptotically~\shortcite{yoshino,Berthier-Bouchaud}.
The crossover is expected at $L_{0}(T) \sim \xi$ and $\tau_{0}(T) \sim
\xi^{z_{eq}}$ with the equilibrium correlation length $\xi =
|T-T_{c}|^{-\nu}$ and $z_{eq}$ and $\nu$ the usual critical exponents
at the critical temperature $T_{c}$.  The values of $T_c$, $\nu$, $z_{eq}$ and
$\theta$ are fixed with equilibrium studies. In the rest of the discussion we 
call $L(t)$ the growing length [$\xi_{\sc sg}(t)\to L(t)$]. 

Similar arguments to the ones exposed for Sinai diffusion and the 
directed manifold imply that the energy density per spin should decay as
$e(t)=e(\infty)+ {\mbox ct}/ L(t)^{d-\theta}$ and provide scaling forms for 
the self-correlation and linear susceptibility of very similar type to the ones 
in previous subsections (but with different values of the exponents).

Numerical tests of the droplet picture predictions were done by a
number of groups. The energy density decay in the $d=3$ and $d=4$ EA
models were checked in~\shortcite{komori} and~\shortcite{yoshino}, respectively.
The numerical spin auto-correlation function $C(t,t_w)$ and the linear
susceptibility $\chi(t,t_w)$ were analyzed along the proposed scaling
forms. In the quasi-equilibrium regime where the FDT
$C(\tau+t_w,t_w)=T\chi(\tau+t_w,t_w)$ holds, the expected scaling
$C(\tau+t_w,t_w)=q_{\sc ea}+{\sc ct} \ T/L(\tau)^{\theta} g_{\sc
  c}(L(\tau)/L(t_w))$ with $g_{\sc c}(\lambda) \propto 1 - {\rm const}
\ \lambda^{d-\theta}+\ldots$ was verified in $d=3$ \shortcite{komori} and
$d=4$ \shortcite{yoshino} (the fact that $L(t)$ is far from a logarithm was
attributed to pre-asymptotics and the analysis was perfomed using the
method sketched above.). Evidence for the validity of droplet aging
scaling for the correlation function $C_{ag}(t,t_w)=f_{\sc
  c}(L(t)/L(t_w))$ and the field cooled susceptibility in which the
sample is cooled in a field that it switched off at $t_w$, $\chi_{\sc
  fc}(t,t_w) =\chi(t,0)-\chi(t,t_w)=L(t)^{-\theta}f_\chi(L(t)/L(t_w))$
with $\chi(t,0)=\chi_{D}-{\rm const}
\ T/L(t)^{\theta}$~\shortcite{Fisher-Huse} in $d=4$ was given
in~\shortcite{yoshino} (note that $\psi\simeq \theta$ is assumed here).
These results were used as support to the standard droplet theory.
However, the asymptotic value of the dynamical susceptibility
$\chi_{\sc D}$ is significantly higher than $\chi_{\sc
  ea}=\beta(1-q_{\sc ea})$~\shortcite{yoshino,rome}. This is clearly at
odds with the standard droplet theory \shortcite{Fisher-Huse} and suggests
the existence of excessive contributions from some unknown soft-modes
other than the droplets in equilibrium.  Moreover, it was found that
the FDT holds up to significantly longer time-differences than
time-translational invariance~\shortcite{yoshino}. These observations
suggest that, even within a droplet perspective, modifications of the
conventional approach are needed.

A different school of thought~\shortcite{rome,JANUS} pushes for a picture
{\it \`a la} mean-field (namely, the dynamics of the SK
model~\shortcite{Cuku2}) that should be accompanied by a complex relaxation
of correlations and susceptibilities in multiple time-scales,
presumably linked to multiple length-scales. So far, this {\it
  ultrametric} organization of time-scales has not been found
numerically~\shortcite{jaubert,JANUS,Bebaku}.

\subsection{Other random systems}

Let us finalize this Section by briefly recalling studies of growing lengths 
in two other systems with quenched randomness: the random ferromagnet
and the sine-Gordon model with random phases.

On the basis of a droplet theory with barriers increasing as 
a power-law of distance, the random ferromagnet should coarsen
with a logarithmically increasing typical length~\shortcite{Huse-Henley}.
Early Monte Carlo simulations~\shortcite{RFM-log} showed agreement with 
this prediction but more recent numerical results from Rieger's group were
interpreted, instead, as evidence for a power-law with variable exponent~\shortcite{Rieger-Paul}.
This result would imply a logarithmic dependence of barriers on the 
domain size. It was argued in~\shortcite{Bustingorry} that the effective power
law maybe the effect of a very long crossover from curvature driven to 
activation with conventional power-law growth of barriers with size. 
Recent simulations support this claim~\shortcite{Park} -- although they
were performed in a dilute ferromagnet.

The Cardy-Ostlund or random sine-Gordon model describes the relaxation dynamics 
of  $2d$ periodic elastic manifolds under the effect of a quenched
random potential. The peculiarity of the model is that 
disorder induced barriers grow only logarithmically with size, $\simeq \ln L$. A 
`super-rough' or `marginal' glassy phase exists below a critical temperature $T_g$. 
The aging dynamics were studied with Coulomb gas and renormalization 
methods close to $T_g$, and with the functional renormalization group (FRG)~\shortcite{Pierre}
and numerical simulations in the full low-$T$ phase~\shortcite{Schehr-CO}. A dynamic 
length $L(t) \sim t^{1/z(T)}$ with $z(T)\simeq 2+c/T$ develops in time, consistently with 
an Arrhenius argument of the kind described in Sect.~\ref{sec:droplet}. 
Intriguingly, $X_\infty=z$ was also found analytically and numerically.

\section{Growing length-scales in aging glasses}
\label{sec:glasses}

In this Section we summarize predictions from mean-field 
theory as well as the outcome of measurements of two-time 
lengths in several glassy systems.

\subsection{Mean-field models}
\label{subsec:glasses}

It has been argued that the Langevin dynamics of mean-field disordered
models are equivalent to mode-coupling theories of glasses. More
precisely, the Sherrington-Kirkpatrick (SK) spin-glass has a dynamic
transition similar to the one found in so-called type A mode-coupling
while $p$-spin models with $p>3$ have a random-first-order transition
(RFOT) similar to the one found in type B mode-coupling
theories~\shortcite{PhysicaA}.  The asymptotic out of equilibrium
relaxation of these models, when the thermodynamic limit has been
taken at the outset, approaches a region of phase space that is not
the one where equilibrium configurations lie.  In models of the RFOT
class this region was named the {\it threshold} since its energy
density is at $O(1)$ distance from the equilibrium
one~\shortcite{Cuku1}. In models of the SK or type A class, although the
configurations visited dynamically are not the ones of equilibrium
their energy density is the same~\shortcite{Cuku2}. In both cases the
region reached dynamically is {\it flat}, in the sense that a dynamic
free-energy landscape can be defined and its geometric properties
studied, and one finds that the eigenvalues of its Hessian mostly
vanish~\shortcite{Cuku1}. The flat directions provide, on the one
hand, channels of aging relaxation and, on the other, their associated
zero-modes give rise to diverging susceptibilities. In the case of
mean-field models the latter cannot be directly associated to
diverging length-scales since these models do not contain any notion
of distance.  Having said this, if one proposes that the same
mechanism, namely the relaxation to a flat region of phase space, is
at the root of aging phenomena in finite dimensional glassy systems,
the diverging susceptibility should be linked to a diverging length
scale. In the super-cooled liquid regime this argument was used by
Franz and Parisi to argue for the existence of a diverging length
scale within the RFOT scenario from the analysis of the $p$-spin
model~\shortcite{Franz-Parisi}, see the chapter by Franz and Semerjian. These ideas were extended to the 
actual MCT in~\shortcite{Biroli-Bouchaud}. In the trully glassy phase a very similar
mechanism is at work and gives support to the symmetry argument~\shortcite{chcu}
that we discuss in Sect.~\ref{sec:reparametrization} and provides a scenario for 
dynamic fluctuations in aging glassy systems.

\subsection{Measurements}
\label{subsec:experiments}

In Sect.~\ref{spinlanguage} we already mentioned simulations~\shortcite{Vollmayr-Lee} and 
experiments~\shortcite{Courtland} in which the 
tagged motion of particles was followed with the intention of characterizing clusters
of more or less mobile ones in aging glasses. 

\begin{figure}[tp]
\centerline{
\includegraphics[width=0.5\textwidth]{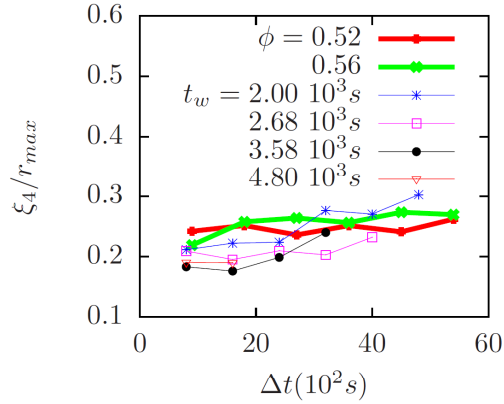}
}
\caption{The two-time dependent growing length in a colloidal
  suspension. The data is taken from~\protect\shortcite{PNAS}. $\phi$ is the
  packing fraction and $r_{max}$ is a cut-off used to compute the
  integrals (see~\protect\shortcite{PNAS} for more details).}
\label{fig-length}
\end{figure}

A two-time dependent dynamic growing length, as defined in Eq.~(\ref{eq:G4}),
was measured with numerical simulations of aging soft spheres and 
Lennard-Jones mixtures~\shortcite{parisi,Parsaeian},
spin-glasses~\shortcite{castillo,jaubert,JANUS} and with confocal microscopy
data of colloidal suspensions~\shortcite{PNAS,Cianci}. An example is 
shown in Fig.~\ref{fig-length} where we display the 
outcome of the data analysis performed in~\shortcite{PNAS}. 

In all cases $\xi_4$ is described by
\begin{eqnarray}
\xi_4(t,t_w) \simeq 
\left\{
\begin{array}{ll}
\xi_{st}(t-t_w)  & \qquad\qquad C>q_{\sc ea}
\nonumber\\
t_w^{a} \ \kappa(C)  & \qquad\qquad C<q_{\sc ea}
\end{array}
\right.
\end{eqnarray}
with $\kappa(C)$ a monotonically decreasing function of $C$
and the exponent $a$ taking a very small value. 
It is important to stress that a clear identification of a finite 
Edwards-Anderson parameter in some of these systems is hard
and the separation of time scales might not be as clear-cut as expressed in
the formula above.

\subsection{Growth processes in kinetically constrained models}

Kinetically constrained models were originally introduced in the 80s
by Fredrickson and Andersen \shortcite{Fredrickson84} as toy models for
understanding glassy dynamics. Generally these are stochastic models
with Markovian dynamics obeying detailed balance with respect to a
(usually) trivial energy function. Some constraints prevent
particular local transitions between configurations~\shortcite{Jackle02}.  
 Despite their trivial equilibrium measure, they
capture many features of real glass-forming systems 
summarized in the chapter by Garrahan, Sollich and Toninelli.

The so-called
spin-facilitated Ising models, are described by the Hamiltonian
$H=-\sum _i n_i$, where $n_i=0,1$ are spin variables on a
$d$-dimensional lattice which can be regarded as a coarse grained
density of moving particles.  The basic idea is that the rearrangement
in a certain region around $i$ will be facilitated if it is surrounded
by a relatively low-density neighbourhood.
In some of these systems, the existence of a growing length in the
non-equilibium evolution after a quench can be easily recognized. Some time 
after a sudden change in temperature towards $T\simeq 0$
all sites are in a dense state with $n_i=1$ and
large frozen regions, the size of which increases as the number of
sites with $n_i=0$ decreases, are present~\shortcite{Crisanti00}.

A particularly interesting
kinetically constrained system is the $2d$ spiral model.
It has the peculiar feature of having a
glass-jamming transition at the critical density
$\langle n_i \rangle=n_c$ of directed percolation~\shortcite{Cristina}.  In
equilibrium above $n_c$ a finite fraction of sites are frozen by the
kinetic constraint.  This means that the configuration space is
partitioned into mutually inaccesible regions; namely, the dynamics is
reducible.  Then all configurations without blocked
regions are dynamically connected to the empty state with $n_i=0 \;
\forall i$, usually denoted as the high-temperature partition, and
they are disconnected from any partition containing frozen regions.
Therefore, quenching the system from (say) the empty state to the
jammed phase, the dynamics do not freeze and remain confined in the
high temperature partition by means of a coarsening process with
a growing correlation length which
resembles in some sense the one observed in binary systems without
kinetic constraints~\shortcite{Corberi09}.

\subsection{Interplay between aging and drive} 
\label{subsec:interplay}

The classical and quantum dynamics of large systems under an external,
non-conservative, drive is a subject of active research. An example in
the first family is the analysis of the rheological properties of soft
glassy materials, that are a result of a competition between the
response to the shearing forces and their intrinsic slow dynamics. An
example in the second class is the study of quantum
magnets under a current generated to the coupling to leads at
different chemical potential. We shall recap here and in Sect.~\ref{subsec:quantum}
a few issues in this
field that are related to the growth of a length and how this is
affected by the drive.

\subsubsection{Flow in coarsening systems}

The effect of an
external drive on coarsening systems is not completely understood and
in many cases a general consensus is lacking. This is at odds with the
wide technological interest of these and similar systems in various
application areas~\shortcite{Larson99}.  In particular, the case of binary
systems in a plane shear flow (Couette flow) has been thoroughly
studied recently.  In this case domains grow elongated and stretched
along the flow direction. This causes ruptures of the
network~\shortcite{Otha90} that may render the segregation incomplete.  The
characteristic lengths in the parallel or perpendicular directions to
the flow could keep on growing indefinitely as without shear or, due
to domain break-up, the system may eventually enter a stationary state
(similarly to what happens in the mean field models discussed above)
characterized by domains with a finite thickness.  In general, there is
no agreement on this point neither from experimental \shortcite{Chan91}
nor theoretical points of view. On the theoretical side, large-$N$
calculations \shortcite{Corberi98} show the existence of an asymptotic
regime with ever growing lengths in all directions.  Stabilization of
the domain size by rupture is excluded also by different approaches
based on approximate theories for scalar systems complemented with RG
analysis~\shortcite{Bray00b}, although with the unexpected feature of
domains growing in the flow direction while shrinking in the
perpendicular one in $d=2$.  Numerical simulations~\shortcite{Padilla97}
cannot still establish a clear-cut evidence due to finite size and
discretization effects. It was also argued \shortcite{Stansell06} that
hydrodynamic effects are crucial in stabilizing a stationary state,
while in the diffusive regime domains might keep on growing.

Another difference with driven mean-field models is provided by the
fluc\-tua\-tion-dissipation relation~$\chi (C)$, the exact computation of
which in the large-$N$ model \shortcite{Corberi02b} excludes a simple
scenario with a single effective temperature. Much more details 
are given in the contribution by Barrat and Lema\^{\i}tre.

\subsubsection{Flow in glassy systems} 

The Langevin dynamics of mean-field disordered models under
non-potential forces has been studied in a series of
papers~\shortcite{Cukulepe,Berthier99}. In all cases (apart from the
spherical model with two-body interactions at zero temperature) aging
is suppressed and a stationary state is reached.  Two-time observables
decay in two steps and the relaxation time decreases with the strength
of the drive as in shear thinning systems.  In the limit of weak drive
the effective temperature takes two values, the one of the bath in the
first relaxation step and a higher value in the second slower stage.
These results have been verified in molecular dynamic simulations of
Lennard-Jones mixtures~\shortcite{Barrat-Berthier}.

The fact that a weak non-potential perturbation can change the
relaxation of a classical system so dramatically, by rendering the
dynamics stationary, can be understood in different ways. First, the
type of non-potential force used in~\shortcite{Cukulepe} can be loosely associated to an infinite temperature noise.
Second, and more importantly, the mechanism for fluctuations based on
time-reparametrization invariance gives a reason for this fact, as
discussed in Sect.~\ref{sec:reparametrization}.

Experimentally, the interplay between aging and flow in an aqueous
suspension of Laponite has been studied by diffusive wave
spectroscopy~\shortcite{Bonn} and light scattering echo
experiments~\shortcite{Pusey} (multiple scattering) and dynamic light
scattering in the single-scattering regime~\shortcite{Ruocco}. Using the
latter technique Di Leonardo {\it et al} measured the density
auto-correlation function and found that as long as the characteristic
relaxation time is smaller than the inverse shear rate aging in
unaffected by the perturbation but, when the relaxation time reaches
the inverse shear rate, aging is strongly reduced.

A recent study of microscopic structural relaxation of 
colloidal suspensions under shear together with relevant references
appeared in~\shortcite{Weeks-recent}. Numerical studies of aging ad plastic 
deformations have been performed by several groups, see the reviews
in~\shortcite{Rottler,JLBarrat}. Nevertherless, as far as we know, an analysis of 
growing dynamic lengths, 
as defined from $G_4$ or $\chi_4$ has not been performed in driven systems 
yet. 

\subsubsection{Aging at the depinning transition of elastic manifolds}

A series of analytic and numerical papers
have recently showed that a non-steady critical 
regime, limited only by the steady correlation length or the system size,
exists  at the zero-temperature  depinning transition 
of an elastic manifold in a disordered medium~\shortcite{Kolton-depinning}. 
The FRG analysis~\shortcite{Pierre} as well as 
molecular dynamics studies prove that there is a driven transient aging regime 
that displays universal features in which a growing length $L(t)\sim t^{1/z}$  
participates in the critical dynamic scaling description of the dynamics.
At the depinning transition the effective height of the 
barriers vanishes justifying the power-law scaling  
the effect of disorder being  a reduction of $z$ to a value smaller
than 2 (at the existent one-loop FRG calculation).

\subsection{Granular matter}
\label{subsec:granular}

The gently driven dynamics of dense granular matter -- athermal
systems -- shows many points in common with the ones of glass forming
liquids~\shortcite{Dauchot}. Aging in a water-saturated granular pile
submitted to discrete taps has been reported in~\shortcite{Debregeas}. The data 
were obtained using multispeckle diffusive wave spectroscopy
to measure particle displacements. Evidence for dynamic heterogeneities and 
a growing correlation length in a horizontally vibrated amorphous assembly 
of hard disks close to the jamming transition was given in~\shortcite{Lechenault}.
Presumably, a two-time growing length would also exist in such 
athermal systems if studied in the aging regime. On the theoretical 
side, the aging properties of a periodically perturbed mean-field model
of the RFOT type and, more precisely, how these depend on the 
amplitude and frequency of the drive were studied in~\shortcite{Becuig}.

\subsection{Quantum fluctuations}
\label{subsec:quantum}

The study of the dynamics of quantum systems has been recently
boosted by the development of cold atom experiments and advanced
measurement techniques.
Quantum magnets are expected to undergo coarsening dynamics in the
ferromagnetic phase. The study of a mean-field model (freely relaxing
and driven by an external current) has revealed an extension of the
super-universality hypothesis to
the quantum realm once small scales are well separated from large
ones, the former feeling the quantum fluctuations and the latter
describing  the motion of large objects -- the walls -- 
being basically identical to the classical ones. More precisely, 
in $O(N)$ field theory  in the large $N$ limit or models of rotors in interaction the length
over which static order is established 
grows as $L(t) \sim t^{1/z}$ with $z=2$ as in the classical limit. This type of coarsening 
survives the application of a voltage difference by two leads at
different chemical potential~\shortcite{Aron}. This prediction is open to experimental 
tests.

\section{A mechanism for dynamic fluctuations}
\label{sec:reparametrization}

In a series of papers Chamon {\it et al} proposed that the symmetry
that captures the universal aging dynamics of glassy systems is the
invariance of an effective dynamical action under {\it uniform
  reparametrizations of the time
  scales}~\shortcite{castillo,jaubert,Chamonetal1,Chamonetal2,Chamonetal3},
see also~\shortcite{Parisi}. 
This approach is reviewed
in~\shortcite{chcu}. Such type of invariance was first encountered in the
{\it mean-field equilibrium dynamics} of
spin-glasses~\shortcite{Sompo,equil-dynamics} and it was later found in the
better formulated {\it out of equilibrium} dynamic of the same
models~\shortcite{Cuku1,Cuku2}.  The invariance means that in the
asymptotic regime of very long times the solution can be found up to a
time-reparametrization transformation.  Within such a scenario,
extended to systems in finite dimensions, the remaining symmetry is
responsible of the dynamic fluctuations.  Support to this claim was
given by the proof of global time reparametrization invariance in the
long times {\it action} of {\it short-range
  spin-glasses}~\shortcite{Chamonetal1,castillo}.  In the proof one
assumes that the dynamics is causal and that there is an additive separation
between a time regime with a fast relaxation and another in which it
is slow [just as in Fig.~\ref{fig-aging} and in Eq.~(\ref{eq:separation})]. The invariance can then be
used to describe dynamic fluctuations in spin-glasses and, as
conjectured, in other glassy systems as well. The fact that the
dynamic action be symmetric under uniform, {\it i.e.} spatially
independent, reparametrizations of time variables [$t\to h(t)$]
suggests that the dynamic fluctuations that cost little action are
those associated to space dependent, long wavelength,
reparametrizations of the form $t\to h({\vec r},t)$. These should be
the Goldstone modes associated with breaking time-reparametrization
invariance symmetry.

In the slow and aging regime the two-time correlation and linear
response depend on both times and time-translation invariance is
lost. As we argued in Sect.~\ref{sec:intro} the two-time dependent
correlation acts as an order parameter.  Dynamic fluctuations are such
that {\it ages} can fluctuate from point to point in the sample with
younger and older pieces (lower and higher values of the correlation)
coexisting at the same values of the two laboratory times: this has
been named {\it heterogeneous aging}.  By looking now at spatially
heterogeneous reparametrizations, we can predict the behavior of local
correlations and linear susceptibilities and the relations between
them.  Different sites can be retarded or advanced with respect to the
global behaviour but they should all have the same overall type of
decay. Similarly, the relation between local susceptibilities and
their associated correlations should be identical all over the
sample~\shortcite{castillo} leading to a uniform effective
temperature~\shortcite{Cukupe}. The reason why the scaling functions should
not fluctuate much is that these are massive modes.  Within the
time-reparametrization invariance scenario the growing correlation
length is due to the approach to the long-time regime in which the
symmetry is fully developped, and the long wavelength modes eventually
become massless.
 
 This hypothesis was put to the test in glassy systems with Monte
 Carlo simulations of the Edwards-Anderson
 model~\shortcite{castillo,jaubert}, molecular dynamics of Lennard-Jones
 mixtures~\shortcite{Parsaeian}, numerical studies of kinetically
 constrained models~\shortcite{Chamonetal2}, and in the study of solvable
 ferromagnetic models with the analysis of the $O(N)$ model in the
 large $N$ limit~\shortcite{Chamonetal3} and the spherical
 ferromagnet~\shortcite{Sollich}.  A series of stringent tests some of them performed in
 these papers were summarized in~\shortcite{chcu}. The outcome of these
 studies is that while trully glassy systems conform to the
 consequences of the hypothesis, simple coarsening as developed in the
 $O(N)$ and spherical ferromagnet does not with time-reparametrization
 invariance being reduced to time rescaling at the heart of the
 difference. This result is very suggestive since it implies that the
 invariance properties and the fluctuations associated to them are
 intimately related to the behavior of the global effective
 temperature (finite against infinite) in the aging regime.

In short, this idea proposes a {\it reason} for the development 
of large dynamic 
fluctuations in glassy dynamics. 

\section{Closing remarks}

In aging systems, the relaxation of typical features in a sample of a
certain age $t_w$ takes a time that increases with $t_w$. Although the
reason for this fact is not necessarily the growth of those features,
the widespread observation of growing lengths in physical systems, as
reviewed in this chapter, promotes this one as the basic mechanism at
the heart of most aging phenomena.  (See~\shortcite{Semerjian2} and the
chapter by Franz and Semerjian for rigurous relations between growing
lengths and growing times.) Related to this fact, the occurrence of
dynamic scaling provides a parametrization of the kinetics in terms of
typical lengths, giving access to universal properties.

Besides these unifying concepts, the very nature of the growing length,
%(and, due to that, the different definitions reviewed in Sec. 2.5),
the growth mechanisms, the relation with equilibrium properties
{\it etc.} are issues which, in most cases, deserve clarification.  In
particular, the interplay between the growth of static order -- as,
in coarsening systems -- and purely dynamical correlations
-- as in most kinetically constrained models -- could
help in understanding the behavior of glasses.  In the case of spin
glasses, moreover, the comprehension of the kind of order established
within the correlated volume is fundamental to distinguish between
competing scenarii, namely droplet-like and {\it \`a la} mean field
pictures.


\begin{thebibliography}{99}

\bibitem{Fisher-Huse} (1) D. S. Fisher and D. A. Huse, 
Phys. Rev. B {\bf 38}, 373 (1988). 

\bibitem{Bouchaud} (2) J.-P. Bouchaud, J. Phys. I {\bf 2}, 1705 (1992).
  J.-P. Bouchaud and D. S. Dean, J. Phys. (France) I {\bf 5}, 265-286
  (1995).
 
\bibitem{Cuku1} (3) L. F. Cugliandolo and J. Kurchan, 
Phys. Rev. Lett. {\bf 71}, 173 (1993).

\bibitem{Cuku2} (4) L. F. Cugliandolo and J. Kurchan,
J. Phys. A {\bf 27}, 5749 (1994). 

\bibitem{Bray94} (5)
A. J. Bray, Adv. Phys. {\bf 43}, 357 (1994).

\bibitem{manifold} (6)
Two reviews on different properties of elastic manifolds that do not, 
however, deal with their aging properties are:
A.-L. Barab\'asi and H. E. Stanley, {\it Fractal Concepts in Surface Growth} 
(Cambridge University Press, Cambridge, 1995).
T. Halpin-Healey and Y.-C. Zhang, Phys. Rep. {\bf 254}, 215 (1995).

\bibitem{Donth}
(7)
 E.  Donth, {\it The glass transition: relaxation
  dynamics in liquids and disordered materials} (Springer, 2001).
  K. Binder and W. Kob, {\it Glassy materials and disordered solids:
    An introduction to their statistical mechanics}, (World
  Scientific, Singapore, 2005).

\bibitem{PNAS} 
(8) C. Chamon, L. F. Cugliandolo, G. Fabricius, J. L. Iguain, 
and E. R. Weeks, PNAS {\bf 105}, 15263 (2008).

\bibitem{chcu} 
(9) C. Chamon and L. F. Cugliandolo, 
J. Stat. Mech. (2007) P07022.

\bibitem{Glotzer} (10)
C. Bennemann, C. Donati, J. Baschnagel, and S.C. Glotzer, Nature (London) 
{\bf 399}, 246 (1999). 
W. Kob, C. Donati, S. J. Plimpton, P. H. Poole, and S. C. Glotzer, Phys. Rev. Lett. {\bf 79},
2827 (1997). 
D. Perera and P. Harrowell, Phys. Rev. E {\bf 54}, 1652 (1996).
 C. Donati et al., Phys. Rev. E {\bf 60}, 3107 (1999).
 A. Onuki and Y. Yamamoto, J. Non-Cryst. Solids {\bf 235-237}, 19 (1998).
 B. Doliwa and A. Heuer, J. Non-Cryst. Solids {\bf 307}, 32 (2002).
 A. Heuer, M. Kunow, M. Vogel, and R.D. Banhatti, Phys. Rev. B {\bf 66}, 224201 
 (2002).

\bibitem{Weitz} (11)
 A. van Blaaderen and P. Wiltzius, Science {\bf 270}, 1177 (1995).
W. K. Kegel and A. van Blaaderen, Science {\bf 287}, 290 (2000).
  E. R. Weeks and D.A. Weitz, Phys. Rev. Lett. {\bf 89}, 095704 (2000).
   E. R. Weeks, J. C. Crocker, A. C. Levitt, A. Schofield, and D. A. Weitz, Science {\bf 287}, 
   627 (2000).
   
\bibitem{Miyagawa} (12)
H. Miyagawa, Y. Hiwatari, B. Bernu, and J-P Hansen,
J. Chem. Phys. {\bf 88}, 3879 (1988). 

\bibitem{Vollmayr-Lee} (13)
K. Vollmayr-Lee, J. Chem. Phys. {\bf 121}, 4781 (2004).
K. Vollmayr-Lee and A. Zippelius, Phys. Rev. E {\bf 72}, 041507 (2005).  
K. Vollmayr-Lee and E. A. Baker, Europhys. Lett. {\bf 76}, 1130 (2006).

\bibitem{Courtland} (14)
R. Courtland and E. R. Weeks, J. Phys.: Cond. Matt. {\bf 15}, S359 (2003).
E. R. Weeks, J. C. Crocker, and D. A. Weitz, J. Phys.: Cond. Matt. {\bf 19}, 205131 (2007).

\bibitem{castillo} (15) H. E. Castillo, C. Chamon, L. F. Cugliandolo, and
  M. P. Kennett, Phys. Rev. Lett. {\bf 88}, 237201 (2002),
  H. E. Castillo, C. Chamon, L. F. Cugliandolo, J. L. Iguain, and
  M. P. Kennett, Phys. Rev. B {\bf 68}, 134442 (2003).

\bibitem{Cukupa} (16) L. F. Cugliandolo, J. Kurchan, and G. Parisi, 
J. Phys. I (France) {\bf 4}, 1641 (1994). 

\bibitem{Lippiello05} (17)
E. Lippiello, F. Corberi, and M. Zannetti, 
Phys. Rev. E {\bf 71}, 036104 (2005). 

\bibitem{Jorge} (18) See, however, the recent proposal in
J. Kurchan and D. Levine, {\it Correlation length for amorphous systems},
 arXiv:0904.4850.

\bibitem{Mepavi} (19) M. M\'ezard, G. Parisi, and M. A. Virasoro, 
{\it Spin glasses and beyond} (World Scientific, 1987).

\bibitem{parisi} (20)
G. Parisi, J. Chem. Phys. B {\bf 103}, 4128 (1999). 

\bibitem{jaubert} (21) L. D. C. Jaubert, C. Chamon, L. F. Cugliandolo, and
  M. Picco, J. Stat. Mech. (2007) P05001.

\bibitem{Semerjian} (22)
G. Semerjian,  L. F. Cugliandolo, and A. Montanari, 
J. Stat. Phys. {\bf 115}, 493 (2004).

\bibitem{Lippiello08} (23)
E. Lippiello, F. Corberi, A. Sarracino, and M. Zannetti, 
Phys. Rev. B {\bf 77}, 212201 (2008);
Phys. Rev. E {\bf 78}, 041120 (2008).

\bibitem{JANUS} (24)
F. Belletti {\it et al},
Phys. Rev. Lett.  {\bf 101}, 157201 (2008); 
J. Stat. Phys. {\bf 135}, 1121 (2009). 

\bibitem{Huse88}  (25)
D. A. Huse, J. Appl. Phys. {\bf 64}, 5776 (1988);
Phys. Rev. B {\bf 43}, 8673 (1991).

\bibitem{Bouchaud05} (26)
J.-P. Bouchaud and G. Biroli, Phys. Rev. B {\bf 72}, 064204 (2005). 

\bibitem{Berthier05} (27)
L. Berthier, G. Biroli, J.-P. Bouchaud, L. Cipelletti, 
D. El Masri, D. L'H\^ote, 
F. Ladieu, and M. Pierno,
Science {\bf 310}, 1797 (2005). 

\bibitem{Caroline} (28)
see, {\it e.g.}
F. Ladieu, C. Thibierge, and D. L'H\^ote, 
J. Phys.: Condensed Matter {\bf 19}, 205138 (2007). 
 C. Thibierge, D. L'H\^ote, F. Ladieu, and R. Tourbot, 
Rev. of Scientific Inst. {\bf 79}, 103905 (2008). 

\bibitem{fluttuazioni} (29)
F. Corberi, E. Lippiello, A. Sarracino, and M. Zannetti,
J. Stat. Mech. P04003 (2010).

\bibitem{Chamonetal1} (30)
C. Chamon, M. P. Kennett, H. E. Castillo, and L. F. Cugliandolo
 Phys. Rev. Lett. {\bf 89}, 217201 (2002). 
 H. E.  Castillo, Phys. Rev. B {\bf  78}, 214430 (2008). 

\bibitem{Chamonetal2} (31)
C. Chamon, P. Charbonneau, L. F. Cugliandolo, D. R. Reichman, and M. Sellitto,
J. Chem. Phys. {\bf 121}, 10120 (2004). 

\bibitem{Chamonetal3} (32)
C. Chamon, L. F. Cugliandolo, and H. Yoshino, 
J. Stat. Mech (2006) P01006.

\bibitem{rieger} (33) H. Rieger, J. Phys. A {\bf  26}, L615 (1993). 
H. Rieger, B. Steckemetz, and M. Schreckenberg, Europhys. Lett. 
{\bf 27}, 485 (1994). J. Kisker, L. Santen, M. Schreckenberg,
and H. Rieger, Phys. Rev. B {\bf 53} 6418 (1996). 

\bibitem{rome} (34) E. Marinari, G. Parisi, F. Ricci-Tersenghi, and
  J. J. Ruiz-Lorenzo, J. Phys. A {\bf 33}, 2373 (2000).

\bibitem{komori} (35) T. Komori, H. Takayama, and H. Yoshino 
  J. Phys. Soc. Japan {\bf 68} 3387 (1999); 
  %Komori T, Takayama H and Yoshino H 1999
  %J. Phys. Soc. Japan 
  {\it ibid} {\bf 69}, 1192 (1999); 
  %Komori T, Takayama H and Yoshino H 2000
  %J. Phys. Soc. Japan 
  {\it ibid} {\bf 69} (Suppl. A) 228 (2000).

\bibitem{yoshino}  (36) H. Yoshino, K. Hukushima, and H. Takayama,
Phys. Rev. B {\bf 66}, 064431 (2002).

\bibitem{Ritort} (37)
A. Crisanti and F. Ritort, 
J. Phys. A {\bf 36}, R181 (2003). 

\bibitem{Cukupe} (38) L. F. Cugliandolo, J. Kurchan, and L. Peliti, 
Phys. Rev. E {\bf 55}, 3898 (1997). 

\bibitem{Godreche02} (39) 
C. Godr\`eche and J. M. Luck, J. Phys. A {\bf 33}, 9141 (2000);
J. Phys.: Condens. Matter {\bf 14}, 1589 (2002).

\bibitem{Janssen89} (40)
H. K. Janssen, B. Schaub, and B. Schmittman, Z. Phys. B Cond.Mat. {\bf 73}, 539 (1989).
P. Calabrese and A. Gambassi, Phys. Rev. E {\bf 65}, 066120 (2002);
Phys. Rev. E {\bf 67}, 36111 (2002);
J. Phys. A: Math. Gen. {\bf 38}, R133 (2005).

\bibitem{Lippiello06} (41) C. Chatelain, J. Stat. Mech. (2004) P06006.
  M. Pleimling and A. Gambassi, Phys. Rev. B {\bf 71}, 180401 (2005).
  P. Mayer, L. Berthier, J. P. Garrahan, and P. Sollich, Phys. Rev.E
  {\bf 68}, 016116 (2005).  E. Lippiello, F. Corberi, and M. Zannetti,
  Phys. Rev. E {\bf 74}, 041113 (2006).  M. Henkel, T. Enss, and
  M. Pleimling, J. Phys. A {\bf 39}, L589 (2006).  F. Corberi,
  A. Gambassi, E. Lippiello, and M. Zannetti, J. Stat. Mech. (2008)
  P02013.

\bibitem{Joubaud09} (42)
S. Joubaud, B. Percier, A. Petrosyan, and S. Ciliberto, Phys. Rev. Lett. {\bf 102}, 130601 (2009).

\bibitem{Bouchaud97} (43)
J.-P. Bouchaud, L. F. Cugliandolo, J. Kurchan, and M. M\'ezard, in {\it Spin Glasses and Random Fields},
edited by A. P. Young (World Scientific, Singapore, 1997).
S. Franz and M. A. Virasoro, J. Phys. A: Math. Gen. {\bf 33}, 891 (2000).

\bibitem{Mazenko04} (44)
G. Mazenko, Phys. Rev. E {\bf 69}, 0116114 (2004).

\bibitem{Lippiello00} (45)
E. Lippiello and M. Zannetti, Phys. Rev. E {\bf 61}, 3369 (2000).

\bibitem{Corberi02} (46)
F. Corberi, E. Lippiello, and M. Zannetti, Phys. Rev. E {\bf 65}, 046136 (2002).

\bibitem{Sicilia} (47)
J. J. Arenzon, A. J. Bray, L. F. Cugliandolo, and A. Sicilia, 
Phys. Rev. Lett. {\bf 98}, 145701 (2007). 
A. Sicilia, J. J. Arenzon, A. J. Bray, and L. F. Cugliandolo,
Phys. Rev. E {\bf 76}, 061116 (2007).

\bibitem{Sicilia-EPL} (48)
A. Sicilia, J. J. Arenzon, A. J. Bray, and L. F. Cugliandolo,
 EPL {\bf 82}, 10001 (2008). 
 
 \bibitem{Henkel-Pleimling} (49)
M. Henkel and M. Pleimling
Phys. Rev. B {\bf 78}, 224419 (2008). 


\bibitem{Simulscaling} (50) F. Corberi, E. Lippiello and M. Zannetti,
  Phys. Rev. E {\bf 68}, 046131 (2003); Phys. Rev. E {\bf 74}, 041106
  (2006).  

\bibitem{Simulscaling2} (51)
C. Aron, C. Chamon, L. F. Cugliandolo, and M. Picco,
  J. Stat. Mech. (2008) P05016.

\bibitem{Puri} (52)
 E. Lippiello, A. Mukherjee, Sanjay Puri, and M. Zannetti, 
 arXiv:1006.0934.
 
\bibitem{Berthier99} (53)
L. Berthier, J.-L. Barrat and J. Kurchan, Eur. Phys. J. B {\bf 11}, 635 (1999).
F. Corberi, E. Lippiello, and M. Zannetti, Phys. Rev. E {\bf 63}, 061506 (2001); 
Eur. Phys. J. B {\bf 24}, 359 (2001).

\bibitem{Chou81} (54)
Y. C. Chou and W. I. Goldburg, Phys. Rev. A {\bf 23}, 858 (1981).
S. Katano and M. Iizumi, Phys. Rev. Lett. {\bf 52}, 835 (1984).
S. Komura, K. Osamura, H. Fujii, and T. Takeda, Phys. Rev.B {\bf 31}, 1278 (1985).
B. D. Gaulin, S. Spooner, and Y. Mori, Phys. Rev. Lett. {\bf 59}, 668 (1987).
A. Sicilia, J. J. Arenzon, I. Dierking, A. J. Bray, L. F. Cugliandolo, J. Mart\'{\i}nez-Perdiguero, 
I. Alonso, and I. C. Pintre, Phys. Rev. Lett. {\bf 101}, 197801 (2008).

\bibitem{Bustingorry} (55)
J. L. Iguain, S. Bustingorry, A. B. Kolton, and L. F. Cugliandolo, 
Phys. Rev. B {\bf 80}, 094201 (2009). 

\bibitem{Bray95} (56)
A. D. Rutenberg and A. J. Bray,  Phys. Rev. Lett. {\bf 74}, 3836 (1995).

\bibitem{Bray00} (57)
A. J. Bray and D. K. Jervis, Phys. Rev. Lett {\bf 84}, 1503 (2000).

\bibitem{Andrenacci06} (58)
N. Andrenacci, F. Corberi, and E. Lippiello, Phys. Rev. E {\bf 74} (2006).

\bibitem{Burioni09} (59)
R. Burioni, F. Corberi, and A. Vezzani, Phys. Rev. E {\bf 79}, 041119 (2009).

\bibitem{Coniglio89} (60)
A. Coniglio and M. Zannetti, Europhys. Lett. {\bf 10}, 575 (1989).
A. Coniglio, P. Ruggiero, and M. Zannetti, Phys. Rev. E {\bf 50}, 1046 (1994).

\bibitem{elastic-line2} (61)
S. Bustingorry, J. L. Iguain, and L. F. Cugliandolo, 
J. Stat. Mech. P09008 (2007). 
Y. L. Chou and M. Pleimling J. Stat. Mech 
P08007 (2010). 

\bibitem{KPZ} (62)
S. Bustingorry, J. Stat. Mech. P10002 (2007).  

%%%%%%%%%%%%%%%%

\bibitem{BM87} (63)
A. J. Bray and M. A. Moore, Phys. Rev. Lett. {\bf 58}, 57 (1987).

\bibitem{Fisher-Huse-static} (64) D. S. Fisher and D. A. Huse, 
Phys. Rev. B {\bf 38}, 386 (1988)

\bibitem{Fisher-Huse-DPRM} (65) D. S. Fisher and D. A. Huse, 
Phys. Rev. B {\bf 43}, 10728 (1991).

\bibitem{vortexglass} (66) D. S. Fisher, M. P. A. Fisher, and D. A. Huse, 
Phys. Rev. B {\bf 43}, 130 (1991). 

\bibitem{Monthus} (67)
C. Monthus and T. Garel, 
J. Stat. Mech. P07002 (2008);
J. Phys. A {\bf 41}, 499801 (2008). 

\bibitem{HY-unpublished} (68) 
H. Yoshino (2001), unpublished.

\bibitem{deCandia} (69)
F. Corberi, A. de Candia, E. Lippiello, and M. Zannetti, Phys. Rev. E {\bf 65}, 046114 (2002).

\bibitem{sinai} (70) Y. G. Sinai, Theor. Probab. Appl.  {\bf 27}, 
247 (1982).

\bibitem{sinai-rsrg} (71) D. S. Fisher, P. Le Doussal, and C. Monthus, 
Phys. Rev. Lett. {\bf 80}, 3539 (1998),
P. Le Doussal, C. Monthus, and D. S. Fisher, 
Phys. Rev. E {\bf 59}, 4975 (1998).

\bibitem{LeDoussal-Vinokur} (72)
%Creep in one dimension and phenomenological theory of glass dynamics
P. Le Doussal and V. M. Vinokur,
Physica C: Superconductivity
{\bf 254}, 63-68 (1995).

\bibitem{Berthier-Bouchaud} (73) L. Berthier and J.-P. Bouchaud, 
Phys. Rev.  B {\bf 66}, 054404 (2002). 

%\bibitem{uppsala-domain-growth} 
%"Domain growth by isothermal aging in 3d Ising and Heisenberg spin glasses"
%P. Jonsson, H. Yoshino, P. Nordblad, H. Aruga Katori and A. Ito,
%Phys. Rev. Lett. {\bf 88}, 257204 (2002).

\bibitem{SY} (74)
%Fragility of the Free-Energy Landscape of a Directed Polymer in Random Media"
M. Sales and H. Yoshino, Phys. Rev. E {\bf 65}, 066131 (2002).

\bibitem{T-chaos-EA} (75)
%Temperature Chaos and Bond Chaos in Edwards-Anderson Ising Spin Glasses: Domain-Wall Free-Energy Measurements
M. Sasaki, K. Hukushima, H. Yoshino, and H. Takayama, 
Phys. Rev. Lett. {\bf 95}, 267203 (2005).

\bibitem{SYM} (76) 
%"A real space renormalization group approach to spin glass dynamics"
F. Scheffler, H. Yoshino, and P. Maass, Phys. Rev. B {\bf 68}, 060404R (2003). 

\bibitem{JYN} (77)
%"Symmetrical Temperature-Chaos Effect with Positive and Negative Temperature Shifts in a Spin Glass"
P. Jonsson, H. Yoshino, and P. Nordblad, Phys. Rev. Lett. {\bf 89}, 097201 (2002)
and 
%Reply to comment on "Symmetrical Temperature-Chaos Effect with Positive and Negative Temperature Shifts in a Spin Glass"
  {\it ibid} {\bf 90}, 059702 (2003).

\bibitem{YLB} (78)
%"Multiple Domain Growth and Memory in the Droplet Model for Spin-Glasses"
H. Yoshino, A. Lemaitre, and J.-P. Bouchaud,  Eur, Phys. J B 
{\bf 20}, 367 (2001). 

\bibitem{MFstep} (79)
%"Step-wise responses in mesoscopic glassy systems: a mean field approach"
H. Yoshino and T. Rizzo Phys. Rev. B {\bf 77}, 104429 (2008).

\bibitem{ghostdomain} (80)
%"Dynamics of ghost domains in spin-glasses"
H. Yoshino, J. Phys. A: Math Gen. {\bf 36} (2003).
%in special issue of Journal of Physics A entitled "Statistical Physics of Disordered Systems: from Real Materials to Optimization and Codes" 

\bibitem{ghoststory} (81)
%Spin Glasses: A Ghost Story
P. E. Jonsson, R. Mathieu, P. Nordblad, 
H. Yoshino, H. Aruga Katori, and A. Ito, Phys. Rev. B. {\bf 70}, 174402 (2004).

\bibitem{PhysicaA} (82)
J-P Bouchaud, L. F. Cugliandolo, J. Kurchan, and M. M\'ezard, 
Physica A {\bf 226}, 243 (1996). 


%%%%%%%%%%%%


\bibitem{Parsaeian} (83)
A. Parsaeian and H. E. Castillo,
Phys. Rev. Lett. {\bf 102}, 055704 (2009). 
Phys. Rev. E {\bf 78}, 060105 (2008). 
H. E. Castillo and A. Parsaeian,
Nature Physics {\bf 3}, 26 (2007). 

\bibitem{Sollich} (84) 
A. Annibale and P. Sollich, J. Stat. Mech. (2009) P02064. 

\bibitem{Berthier} (85)
L. Berthier, Phys. Rev. Lett. {\bf 98}, 220601 (2007).

\bibitem{elastic-line} (86)
A. Barrat, Phys. Rev. E {\bf 55}, 5651 (1997).
H. Yoshino, J. Phys. A {\bf 29}, 
1421 (1996); Phys. Rev. Lett. {\bf 81}, 1493 (1998).
S. Bustingorry, J. L. Iguain, C. Chamon, L. F. Cugliandolo,
and D. Dom\'{\i}nguez, 
Europhys. Lett. {\bf 76}, 856 (2006).

\bibitem{Bebaku} (87)
L. Berthier, J.-L. Barrat, and J. Kurchan, 
Phys. Rev. E {\bf 63}, 016105 (2001).

%%%%%%%%%% Random ferroma

\bibitem{Huse-Henley} (88)
D. A. Huse and C. L. Henley, 
Phys. Rev. Lett. {\bf 54}, 2708 (1985). 

\bibitem{RFM-log} (89)
S. Puri, D. Chowdhury, and N. Parekh, J. Phys. A {\bf 24}, L1087 (1991).
A. J. Bray and K. Humayun, J. Phys. A {\bf 24}, L1185 (1991). 

\bibitem{Rieger-Paul} (90)
R. Paul, S. Puri, and H. Rieger, 
Europhys. Lett. {\bf 68}, 881 (2004);
Phys, Rev. E {\bf 71}, 061109 (2005).
 H. Rieger, G. Schehr, and 
R. Paul, Prog. Thoer. Phys. Suppl. {\bf 157}, 111 (2005). 

\bibitem{Park} (91)
H. Park and M. Pleimling, 
 arXiv:1009.1677.

%%%%%%% FRG Cardy-Ostlund
 \bibitem{Pierre} (92)
 P. Le Doussal, 
{\it Exact results and open questions in first principle functional RG},
  arXiv:0809.1192.

 \bibitem{Schehr-CO} (93)
 G. Schehr and P. Le Doussal, Phys. Rev. E {\bf 68}, 046101 (2003); 
 Phys. Rev. Lett. {\bf 93}, 217201 (2004). 
 G. Schehr and H. Rieger, Phys. Rev. B {\bf 71}, 184202 (2005).
 
%%%%%%%%%%%%%%%

%%%%%%%%%%%%%%

\bibitem{Franz-Parisi}  (94)
S. Franz and G. Parisi, J. Phys. C {\bf 12}, 6335 (2000).

\bibitem{Biroli-Bouchaud} (95)
G. Biroli and J-P Bouchaud, 
Europhys. Lett. {\bf 67}, 21 (2004). 

\bibitem{Cianci} (96)
G. C. Cianci, R. E. Courtland, and E. R. Weeks,
Solid State Comm. {\bf 139}, 599 (2006).
P. Yunker, Z. X.  Zhang, K. B. Aptowicz, and A. G. Yodh
Phys. Rev. Lett. {\bf 103},  115701 (2009). 

\bibitem{Fredrickson84} (97)
G. H. Fredrickson and H. C. Andersen, Phys. Rev. Lett. {\bf 53}, 1244 (1984);
J. Chem.  Phys. {\bf 83}, 5822 (1985).

\bibitem{Jackle02} (98)
J. J\"ackle, J. Phys. Cond.Matt. {\bf 14}, 1423 (2002).
P. Sollich and F. Ritort, Adv. in Phys. {\bf 52}, 219 (2003).
S. Leonard, P. Mayer, P. Sollich, L. Berthier,  and J. P. Garrahan, J. Stat. Mech. (2007) 
P07017.

\bibitem{Crisanti00} (99)
A. Crisanti, F. Ritort, A. Rocco,  and M. Sellitto, J. Chem. Phys. {\bf 113}, 10615 (2000).
P. Sollich and M. R. Evans, Phys. Rev. Lett. {\bf 83}, 3238 (1999).
J. P. Garrahan and M. E. J. Newman, Phys. Rev. E {\bf 62}, 7670 (2000).
S. N. Majumdar, D. S. Dean,  and P. Grassberger, Phys. Rev. Lett. {\bf 86}, 2301 (2001).

\bibitem{Cristina} (100) C. Toninelli and G. Biroli, 
G. Biroli and C. Toninelli, Eur. Phys. J. B {\bf 64} 567 (2008); 
C. Toninelli and G. Biroli, J. Stat. Phys. 130 83 (2008).

\bibitem{Corberi09} (101)
F. Corberi and L. F. Cugliandolo, 
J. Stat. Mech. (2009) P09015.

%%%%%%%%%%%%%%%%

\bibitem{Larson99} (102)
See, {\it e.g.} R. G. Larson, 
{\it The structure and Rheology of Complex fluids} (Oxford University Press, New York, 1999).

\bibitem{Otha90} (103)
T. Otha, N. Nozaki and M. Doi, Phys. Lett. A {\bf 145}, 304 (1990);
J. Chem. Phys. {\bf 93}, 2664 (1991).

\bibitem{Chan91} (104)
C. K. Chan, F. Perrot,  and D. Beysens, Phys. Rev. A {\bf 43}, 1826 (1991).
T. Takebe, F. Fujioka, R. Sawaoka,  and T. Hashimoto, J. Chem. Phys. {\bf 98}, 717 (1993).
T. Hashimoto, K. Matsuzaka, E. Moses,  and A. Onuki, Phys. Rev. Lett. {\bf 74}, 126 (1995).
J.  L\"auger, C. Laubner,  and W. Gronsky, Phys. Rev. Lett. {\bf 75}, 3576 (1995).
R. Yamamoto and X. C. Zeng, Phys. Rev. E {\bf 59}, 3223 (1997).
Z. Shou and A. Chakrabarti, Phys. Rev. E {\bf 61}, R2200 (2000).
H. Liu and A. Chakrabarti, J. Chem. Phys. {\bf 112}, 10582 (2000).

\bibitem{Corberi98} (105)
F. Corberi, G. Gonnella, and A. Lamura, Phys. Rev. Lett. {\bf 81}, 3852 (1998);
Phys. Rev. E {\bf 61}, 6621 (2000).
N. P. Rapapa and A. J. Bray, Phys. Rev. Lett. {\bf 83}, 3856 (1999).

\bibitem{Bray00b} (106)
A. J. Bray and A. Cavagna, J. Phys. A {\bf 33}, L305 (2000).
A. J. Bray, A. Cavagna, and Rui D. M. Travasso, Phys. Rev. E {\bf 64}, 012102 (2001);
Phys. Rev. E {\bf 65}, 016104 (2002).

\bibitem{Padilla97} (107)
P. Padilla and S. Toxvaerd, J. Chem. Phys. {\bf 106}, 2342 (1997).
M. E. Cates, V. M. Kendon, P. Bladon, and J.-C. Despat, Faraday Discuss. {\bf 112}, 1 (1999).
F. Corberi, G. Gonnella, and A. Lamura, Phys. Rev. Lett. {\bf 83}, 4057 (1999);
Phys. Rev. E {\bf 62}, 8064 (2000). 

\bibitem{Stansell06} (108)
P. Stansell, K. Stratford, J.-C.Despat, R. Adhikari and M. E. Cates, 
Phys. Rev. Lett. {\bf 96}, 085701 (2006).

\bibitem{Corberi02b} (109)
F. Corberi, G. Gonnella, E. Lippiello and M. Zannetti, Europhys. Lett. {\bf 60}, 425 (2002);
J. Phys. A {\bf 36}, 4729 (2003).


%%%%%%%%%%%%%%%%
\bibitem{Cukulepe} (110)
L. F. Cugliandolo, J. Kurchan, P. Le Doussal, and L. Peliti,
Phys. Rev. Lett. {\bf 78}, 350 (1997).

\bibitem{Barrat-Berthier} (111)
J-L Barrat and L. Berthier, Phys. Rev. Lett. {\bf 89},  095702 (2002). 
J. Chem. Phys. {\bf 116},  6228 (2002). 


\bibitem{Bonn} (112)
D. Bonn, S. Tanase, B. Abou, H. Tanaka, and J. Meunier, 
Phys. Rev. Lett. {\bf 89}, 015701 (2002). 
 B. Viasnoff and F. Lequeux, Phys. Rev. Lett. 
{\bf 89}, 065701 (2002). 

\bibitem{Pusey} (113)
G. Petekis, A. Moussald, and P. N. Pusey, 
Phys. Rev. E {\bf 66}, 051402 (2002). 

\bibitem{Ruocco} (114)
%   - "Aging under shear: structural relaxation of a
%   non-Newtonian fluid.
   R. Di Leonardo, F. Ianni, and G. Ruocco,
   Phys. Rev. E {\bf 71}, 011505 (2005).

 \bibitem{Weeks-recent}  (115)
D. Chen, D. Semwogerere, J. Sato,
   V. Breedveld, and E. R.  Weeks, arXiv:0908.4226.

\bibitem{Rottler} (116)
J. Rottler and M. Warren, Eur. Phys. J. Sp. Topics {\bf 161}, 
55 (2008). 

\bibitem{JLBarrat} (117)
J-L Barrat, J. Baschnagel, and A. Lyulin, 
arXiv:1002.2065.


%%%%%%%%%%%%%% Depinning 

 \bibitem{Kolton-depinning} (118)
 G. Schehr and P. Le Doussal, Europhys. Lett. {\bf 71}, 290 (2005).
 A. B. Kolton, A. Rosso, E. V. Albano, and T. Giamarchi,
Phys. Rev. B {\bf 74}, 140201(R) (2006).
 A. B. Kolton, G. Schehr, and P. Le Doussal, 
 Phys. Rev. Lett. {\bf 103}, 160602 (2009).
 

%%%%%%%%%%%%%%% Granular

\bibitem{Dauchot} (119) 
O. Dauchot,  {\it Glassy behaviours in a-thermal systems, 
the case of granular media: A tentative review},
Lecture Notes in Physics {\bf 716}, 161 (2007).

\bibitem{Debregeas} (120) 
A. Kabla and G. Debregeas,
Phys. Rev. Lett. {\bf 92}, 035501 (2004). 

\bibitem{Lechenault} (121)
O. Dauchot, G. Marty, and G. Biroli,
Phys. Rev. Lett. {\bf 95}, 265701 (2005). 
F. Lechenault, O. Dauchot, G. Biroli, and J-P Bouchaud, 
EPL {\bf 83},  46002 (2008).

\bibitem{Becuig} (122)
L. Berthier, L. F. Cugliandolo, and J. L. Iguain, 
Phys. Rev. E {\bf 63}, 051302 (2001). 

\bibitem{Aron} (123) C. Aron, G. Biroli, and L. F. Cugliandolo, 
Phys. Rev. Lett. {\bf 102}, 050404 (2009). 

\bibitem{Parisi} 
 S. Franz, G. Parisi, F. Ricci-Tersenghi, and T. Rizzo, 
arXiv:1008.0996.

\bibitem{Sompo} (124) 
H. Sompolinsky, Phys. Rev. Lett. {\bf 47}, 935 (1981).

\bibitem{equil-dynamics}  (125) 
S. L. Ginzburg, Zh. Eksp. Teor. Fiz. {\bf 90}, 754
(1986) [Sov. Phys. JETP {\bf 63}, 439 (1986)].
L. B. Ioffe, Phys. Rev. B {\bf 38}, 5181 (1988).

\bibitem{Semerjian2} (126)
A. Montanari and G. Semerjian,
J. Stat. Phys. {\bf 125}, 23 (2006).
 
\end{thebibliography}
\end{document}